\documentclass{elsarticle}

\usepackage{hyperref}
\usepackage{textcomp}
\usepackage{multirow}
\usepackage{floatrow}
\usepackage{amssymb} % Need amssymb for \lesssim

\usepackage{natbib}
\newcommand{\citen}[1]{\citet{#1}}
\renewcommand{\cite}[1]{\citep{#1}}
\def\Fig#1{Figure~\ref{#1}}
\def\Tab#1{Table~\ref{#1}}
\def\Sect#1{Section~\ref{#1}}

\journal{Icarus}

\biboptions{authoryear}
\begin{document}

\begin{frontmatter}

\title{Thermal Properties and an Improved Shape Model for Near-Earth Asteroid (162421) 2000 ET70}

\author[cornell]{Sean E. Marshall\corref{mycorrespondingauthor}}
\cortext[mycorrespondingauthor]{Corresponding author at: 525 Space Sciences Building, Cornell University, Ithaca, NY 14853, USA}
\ead[url]{astro.cornell.edu/~seanm/}
\ead[email]{seanm@astro.cornell.edu}

\author[lpl]{Ellen S. Howell\fnref{irtf_visit}}
\author[maine]{Christopher Magri}
\author[apl]{Ronald J. Vervack, Jr.\fnref{irtf_visit}}
\author[cornell]{Donald B. Campbell}
\author[ucf]{Yanga R. Fern{\'a}ndez\fnref{irtf_visit}}
\author[lpl]{Michael C. Nolan}
\author[ucf]{Jenna L. Crowell}
\author[jpl]{Michael D. Hicks}
\author[jpl]{Kenneth J. Lawrence}
\author[arecibo]{Patrick A. Taylor}

\fntext[irtf_visit]{Visiting Astronomers at the InfraRed Telescope Facility, which is operated by the University of Hawaii under contract NNH14CK55B with the National Aeronautics and Space Administration.}
% The line from http://irtfweb.ifa.hawaii.edu/research/acknowledge.php
\address[cornell]{Department of Astronomy, Cornell University, Ithaca, NY 14853, USA}
\address[lpl]{Lunar and Planetary Laboratory, University of Arizona, Tucson, AZ 85721, USA}
\address[maine]{University of Maine at Farmington, Farmington, ME 04938, USA}
\address[apl]{Johns Hopkins University / Applied Physics Laboratory, Laurel, MD 20723, USA}
\address[ucf]{Department of Physics, University of Central Florida, Orlando, FL 32828, USA}
\address[jpl]{Jet Propulsion Laboratory, California Institute of Technology, Pasadena, CA 91109, USA}
\address[arecibo]{Arecibo Observatory, Universities Space Research Association, Arecibo, PR 00612, USA}

\begin{abstract}
We present thermal properties and an improved shape model for potentially hazardous asteroid (162421) 2000~ET70. In addition to the radar data from 2000~ET70's apparition in 2012, our model incorporates optical lightcurves and infrared spectra that were not included in the analysis of Naidu et al. (2013, \textit{Icarus} 226, 323-335). We confirm the general ``clenched fist'' appearance of the Naidu et al. model, but compared to their model, our best-fit model is about 10\% longer along its long principal axis, nearly identical along the intermediate axis, and about 25\% shorter along the short axis. We find the asteroid's dimensions to be 2.9~km~\texttimes~2.2~km~\texttimes~1.5~km (with relative uncertainties of about 10\%, 15\%, and 25\%, respectively).
With the available data, 2000~ET70's period and pole position are degenerate with each other. The radar and lightcurve data together constrain the pole direction to fall along an arc that is about twenty-three~degrees long and eight~degrees wide.
Infrared spectra from the NASA InfraRed Telescope Facility (IRTF) provide an additional constraint on the pole. Thermophysical modeling, using our SHERMAN software, shows that only a subset of the pole directions, about twelve~degrees of that arc, are compatible with the infrared data. Using all of the available data, we find that 2000~ET70 has a sidereal rotation period of 8.944~hours~(\textpm~0.009~h) and a north pole direction of ecliptic coordinates $(52^{\circ}, -60^{\circ}) \pm 6^{\circ}$.
The infrared data, acquired over several dates, require that the thermal properties (albedo, thermal inertia, surface roughness) must change across the asteroid's surface. By incorporating the detailed shape model and spin state into our thermal modeling, the multiple ground-based observations at different viewing geometries have allowed us to constrain the levels of the variations in the surface properties of this asteroid.
\end{abstract}


\end{frontmatter}

%\linenumbers

\section{Introduction to 2000~ET70 \label{sec_intro}}

Near-Earth asteroid (162421)~2000~ET70 (hereafter ET70) was discovered on March~8,~2000, by the Lincoln Near-Earth Asteroid Research (LINEAR) program in Socorro, New~Mexico. It is an Aten asteroid ($a =$~0.947~au).
\citen{Williams2000} and \citen{Whiteley2001} reported an absolute visible magnitude of 18.2, but that value was based on observations at a phase angle of 70~degrees.

ET70 passed near Earth in February~of~2012, with a closest approach distance of 0.0454~au (18~lunar distances) on February~19. A series of radar observations with the 305-meter William E. Gordon Telescope at the Arecibo Observatory and with the 70-meter DSS-14 antenna at NASA's Goldstone Deep Space Communications Complex provided continuous-wave spectra and delay-Doppler images that yielded a shape model \cite{Naidu2013}, somewhat reminiscent of a clenched fist -- roughly ellipsoidal with ridges and valleys near its north pole. Naidu et al. found ET70's dimensions to be 2.61~km~$\times$~2.22~km~$\times$~2.04~km (with uncertainties of 5\%). ET70's size and its Minimum Orbit Intersection Distance (MOID) with respect to Earth of 0.03~au make it a potentially hazardous asteroid (PHA).

Using the \citen{Naidu2013} shape, we attempted to model the thermal emission from ET70 as constrained by spectra we obtained at NASA's InfraRed Telescope Facility (IRTF). Our early thermal modeling runs suggested that the pole might be further south than the position at ecliptic coordinates $(\lambda, \beta) = (80^{\circ}, -50^{\circ}) \pm 10^{\circ}$ obtained by Naidu et al. This motivated a reassessment of their shape model, which was based solely upon the radar observations.

Lightcurve observations of ET70 that were also obtained in February~2012 can provide additional constraints for the shape modeling process, particularly the determination of the pole location. We have therefore revisited the Naidu et al. model using both the radar and lightcurve data in concert with thermal modeling of our IR observations in order to improve the shape model for ET70, with a focus on the pole position, and to determine the best estimates of its thermal parameters based on its revised shape. This study demonstrates the power of multiple data sets in the investigation of near-Earth asteroids (NEAs).

\section{Observations \label{sec_obs}}

\subsection{Radar observations \label{sec_obs_radar}}

Radar provides by far the best way to spatially resolve NEAs from the ground, by using observations in which a powerful series of radio waves is transmitted toward an asteroid and the echoes that reflect off the asteroid are received after the round-trip light travel time. The echoes can be analyzed in time (delay) and frequency (Doppler shift) to produce two-dimensional delay-Doppler radar images of the asteroid, a technique that has also been applied to map other planetary bodies \citep[e.g.][]{Ostro1993, Campbell2006}.

Each pixel in a delay-Doppler image includes the contributions from all parts of the target's surface that have the same distance and line-of-sight velocity relative to the observer. For a convex object, most delay-Doppler pixels include contributions from two different locations on the surface -- leading to the so-called north-south ambiguity -- whereas more complicated shapes can have three or more locations contributing to some of the delay-Doppler pixels \cite{Ostro2002}. Careful analysis of radar images acquired over the course of an asteroid's rotation makes it possible to determine the asteroid's three-dimensional shape, size, and rotation state, often with great accuracy \citep[e.g.][]{Magri2007, Magri2011, Nolan2013}. Surface resolutions of a few meters are sometimes possible, much better than can be achieved by any other Earth-based technique.

A second type of radar observation is a continuous wave (CW) radar spectrum. In these one-dimensional spectra, the target is resolved in frequency but not in time. These Doppler spectra therefore have a higher signal-to-noise ratio (SNR) in each frequency bin than simultaneous delay-Doppler images would have in each pixel. CW spectra are the observing mode of choice when the target is expected to be relatively faint, due to a large distance from the observer.

All of the radar observations of ET70 took place in 2012. We have delay-Doppler images and CW spectra from ten days between February~12 and February~23, 2012, around the time of ET70's closest approach (0.05~au). There were additional radar observations on two days in August~2012 but, due to ET70's greater distance (0.16~au) at that time, only CW spectra could be acquired. The August CW spectra were used in shape modeling, but they were not particularly helpful in constraining ET70's properties. The details of the radar observations are given in \Tab{tab_obs_radar}. We are using all of the radar data that were used by \citen{Naidu2013}, but we have summed some of the images differently. We also have incorporated some lower-quality radar data sets that were not used for the final shape model of Naidu et al.: a few coarse-resolution delay-Doppler images from Goldstone and some additional CW spectra from Arecibo.

\begin{table}[!hb]
\begin{tabular}{ccccrrrrrc}
\hline
UT date & UT times & Tel & Mode & $\delta r$ (m) & $\delta f$ (Hz) & Runs & RTT (s) & $P_{tx}$ (kW) \\\hline
% Date &     UT times    & Tel & mo & dr &   df  &  Runs   & RTT & Ptx
02-12 & 08:27:51--08:37:55 & A & CW &    & 0.167 &   5     &  67 & 828 \\
      & 08:42:47--10:29:47 &   & DD & 15 & 0.075 &  48     &     &     \\
      & 10:53:18--11:07:41 &   & DD & 15 & 0.075 &   7     &     &     \\\hline
02-13 & 08:11:06--08:25:56 & A & CW &    & 0.182 &   7     &  62 & 860 \\
      & 08:30:34--10:53:26 &   & DD & 15 & 0.075 &  50     &     &     \\\hline
02-14 & 07:59:56--08:04:43 & A & CW &    & 0.196 &   3     &  58 & 811 \\
      & 08:06:40--10:19:45 &   & DD & 15 & 0.075 &  59     &     &     \\\hline
02-15 & 07:53:54--08:00:11 & A & CW &    & 0.213 &   4     &  54 & 785 \\
      & 08:03:01--08:14:46 &   & DD & 15 & 0.075 &   5     &     &     \\
      & 08:18:28--10:09:46 &   & DD & 15 & 0.075 &  58     &     &     \\\hline
02-15 & 09:17:49--09:33:20 & G & DD & 75 & 1.532 &   9     &  54 & 420 \\
      & 09:46:26--12:24:09 &   & DD & 37 & 1.021 &  77\dag &     &     \\\hline
02-16 & 07:34:18--07:38:30 & A & CW &    & 0.227 &   3     &  51 & 760 \\
      & 07:48:38--07:51:06 &   & CW &    & 0.227 &   2     &     &     \\
      & 07:53:28--09:36:34 &   & DD & 15 & 0.075 &  61     &     &     \\\hline
02-16 & 12:15:56--13:28:09 & G & DD & 15 & 1.000 &  29\dag &  51 & 420 \\
      & 13:29:06--15:29:31 &   & DD & 15 & 1.000 &  70\dag &     &     \\\hline
02-17 & 07:38:00--07:41:57 & A & CW &    & 0.244 &   3     &  48 & 775 \\
      & 07:46:14--08:48:59 &   & DD & 15 & 0.075 &  39     &     &     \\\hline
02-17 & 07:42:56--08:00:01 & G & DD & 75 & 1.532 &  11\dag &  48 & 420 \\
      & 08:16:58--12:24:19 &   & DD & 37 & 0.977 & 152\dag &     &     \\\hline
02-18 & 07:36:05--07:50:52 & G & DD & 75 & 1.532 &  10\dag &  47 & 420 \\
      & 08:01:16--08:45:24 &   & DD & 37 & 0.977 &  26     &     &     \\\hline
02-19 & 07:21:56--07:36:25 & G & DD & 75 & 1.532 &  10\dag &  46 & 420 \\
      & 07:46:13--13:07:48 &   & DD & 37 & 0.977 & 188\dag &     &     \\\hline
02-20 & 08:12:15--11:26:19 & G & DD & 37 & 0.977 &  80\dag &  46 & 420 \\\hline
02-23 & 09:20:47--10:55:20 & G & DD & 75 & 0.977 &  55\dag &  51 & 420 \\\hline
08-24 & 15:46:51--16:31:17 & A & CW &    & 0.342 &   9     & 153 & 721 \\\hline
08-26 & 15:04:24--16:15:22 & A & CW &    & 0.333 &  14     & 157 & 722 \\\hline
\end{tabular}
\caption{
Radar observations used in this work. All observations are from 2012.
% Between February 12 and 17, sky motion of ET70 was about 5.49 deg/day
% MJD was removed
UT times specify when data reception began and ended.
The third column specifies which telescope was used for the observations. The Arecibo planetary radar (A) uses S-band (2380~MHz frequency, 12.6~cm wavelength), and the Goldstone Solar System Radar (G) uses X-band (8560~MHz, 3.5~cm).
The fourth column (mode) specifies whether the observations were continuous-wave spectra (CW) or delay-Doppler images (DD).
For delay-Doppler images, $\delta r$, the range resolution, specifies the equivalent distance from the time delay resolution of the data. For instance, delay resolution of 0.1~microseconds (100~nanoseconds) is equivalent to a range resolution of 15~meters.
% CW spectra measure frequency only -- no delay measurements.
$\delta f$ specifies the frequency resolution of the data that were used in the shape fitting.
The numbers of runs listed here are the numbers of usable scans (the numbers of round-trip times for which data was acquired) for each data set. A dagger (\dag) indicates that subsets of these data were summed before being input to the shape fitting software, in order to increase SNR. For instance, the 77 Goldstone imaging scans on February~15 were summed, with (most) sums having three scans, so that the shape modeling software used 26 sums.
RTT is the round-trip time to the asteroid for that set of observations. A round-trip time of 60~seconds corresponds to a distance of 0.060~au.
$P_{tx}$ is the transmitter power.
}
\label{tab_obs_radar}
\end{table}

\subsection{Lightcurve observations \label{sec_obs_lc}}

\citen{Alvarez2012} obtained lightcurve observations of ET70 from February~19 through 24, 2012, from four different locations (see \Tab{tab_obs_lightcurve}), and submitted them to the Minor Planet Center's Asteroid Light Curve Database \cite{Warner2011}. \citen{Alvarez2012} found a rotation period for ET70 of $8.947\pm0.001$~hours. \citen{Naidu2013} noted that without information on ET70's pole position, this observed period allows for sidereal rotation periods from 8.902~to~8.992~hours.

\begin{table}
\begin{tabular}{lcrcccllrr}
\hline
\multirow{2}{*}{UT date} & \multirow{2}{*}{UT times} & \multirow{2}{*}{$\Delta\phi$} & $r$  & $\Delta$ & \multirow{2}{*}{$\alpha$} & Observatory & \multirow{2}{*}{Observer} & \multicolumn{2}{c}{Data points} \\
{}                       & {}                        & {}                            & (au) &   (au)   &                {}      & (and MPC code) & {}                        & Raw & Dec. \\\hline
02-19 & 03:31--08:54 & $217^{\circ}$ & 1.020 & 0.046 & $45^{\circ}$ & OLASU (I38)      & Alvarez &  487 & 21 \\
02-21 & 05:53--13:05 & $290^{\circ}$ & 1.024 & 0.047 & $41^{\circ}$ & Kitt Peak (695)  & Han     & 1027 & 42 \\
02-22 & 01:37--07:10 & $224^{\circ}$ & 1.025 & 0.048 & $41^{\circ}$ & OLASU            & Alvarez &  531 & 18 \\
02-22 & 09:45--14:24 & $187^{\circ}$ & 1.025 & 0.049 & $41^{\circ}$ & Kingsgrove (E19) & Oey     &  235 & 24 \\
02-23*& 23:59--05:47 & $234^{\circ}$ & 1.026 & 0.050 & $42^{\circ}$ & OLASU            & Alvarez &  582 & 18 \\
02-23 & 09:35--18:20 & $352^{\circ}$ & 1.027 & 0.052 & $43^{\circ}$ & Kingsgrove       & Oey     &  377 & 35 \\
02-24 & 00:15--01:30 & $ 50^{\circ}$ & 1.028 & 0.053 & $43^{\circ}$ & Cerro Tololo     & Han     &  120 &  9 \\
02-24 & 09:36--18:38 & $364^{\circ}$ & 1.028 & 0.055 & $44^{\circ}$ & Kingsgrove       & Oey     &  264 & 33 \\\hline
\end{tabular}
\caption{
%Lightcurve observations used in this work. All of these observations were taken within a few days of the New Moon of February 21, 2012\footnote{\url{http://aa.usno.navy.mil/cgi-bin/aa_phases.pl?year=2012&month=1&day=1&nump=50&format=p}}.
Lightcurve observations used in this work. All of these observations were taken within a few days of the \href{http://aa.usno.navy.mil/cgi-bin/aa_phases.pl?year=2012&month=1&day=1&nump=50&format=p}{New Moon} of February 21, 2012.
%\footnotemark.
%Note that the February~21 lightcurve from Kitt Peak was not listed in Table~1 of \citen{Alvarez2012}, though it was uploaded to the Minor Planet Center\footnote{\url{http://alcdef.org/alcdef_GenerateALCDEFPage.php}}.
Note that the February~21 lightcurve from Kitt Peak was not listed in Table~1 of \citen{Alvarez2012}, though it was uploaded to the Minor Planet Center (\href{http://alcdef.org/alcdef_GenerateALCDEFPage.php}{ALCDEF}).
%\footnotemark.
$\Delta\phi$ is the range of asteroid rotation phases covered by that night's lightcurves, assuming the best-fit sidereal rotation period of 8.944~hours. A value greater than $360^{\circ}$ indicates that the lightcurves encompassed more than a full rotation of ET70.
$r$ is the distance from the asteroid to the Sun; $\Delta$ is the distance from the asteroid to the Earth; $\alpha$ is the solar phase angle.
The final two columns list the numbers of data points in the raw and decimated versions of each lightcurve.\newline
* First observation was at 23:59~UT on February~22; last observation was at 05:47~UT on February~23
}
\label{tab_obs_lightcurve}
%\footnotetext{\url{http://aa.usno.navy.mil/cgi-bin/aa_phases.pl?year=2012&month=1&day=1&nump=50&format=p}}
%\footnotetext{\url{http://alcdef.org/alcdef_GenerateALCDEFPage.php}}
\end{table}

\begin{figure}
\begin{center}
\includegraphics[width=0.9\textwidth]{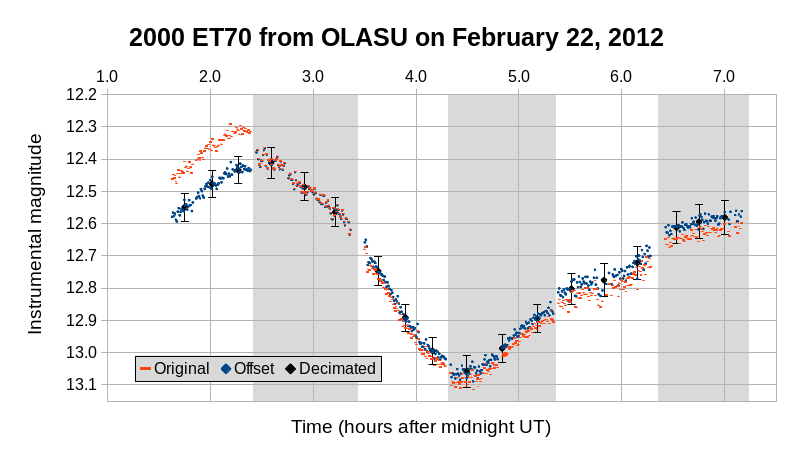} %{Feb22a_OLASU_paper3.png}
% https://en.wikibooks.org/wiki/LaTeX/Importing_Graphics
\end{center}
\caption{
Example of lightcurve data at different stages of processing. The alternating white and gray backgrounds indicate the different sessions. Because each session had a different set of comparison stars, there are offsets between the magnitudes from different sessions. The magnitudes from each session had to be shifted vertically, so that the combined lightcurve's amplitude would be consistent with the amplitude of the composite lightcurve in \citen{Alvarez2012}. The orange points show the original data (with magnitude offsets not corrected), the blue points show the combined lightcurve (with appropriate magnitude offsets for the different sessions), and the black points with error bars show the final combined and decimated lightcurve that was used for shape modeling. Typically, the decimated lightcurves had three points from each session.
There were two contributions to the error bars of the decimated lightcurves: uncertainty in the offsets between the different sessions (0.03 to 0.05~magnitudes) and uncertainty from the scatter in the original (undecimated) lightcurves' data points (usually 0.01 to 0.03~mag).
}
\label{fig_lc_Feb22a_data}
\end{figure}

Because the asteroid was moving fairly quickly across the sky (about $7^{\circ}$ per day during the lightcurve observations), each night's observations had to be divided into multiple sessions, with each session having a different set of comparison stars. (See Supplementary online Figure~S1 for a plot of ET70's sky positions during all measurements.) Guided by the composite lightcurve in \citen{Alvarez2012}, we joined the segments from the different sessions, and combined them into a single consistent lightcurve for each night, an example of which is shown in \Fig{fig_lc_Feb22a_data}. In order to speed up the computations, all lightcurves were decimated in time before being input to the shape modeling software.

\begin{figure}
\floatbox[{\capbeside\thisfloatsetup{capbesideposition={right,center}}}]{figure}[\FBwidth]
{\caption{
Lightcurve observations from Kingsgrove on February~24, 2012 (black points with error bars; decimated). Note the asymmetric maxima, which were a key feature of multiple lightcurves. The blue curve shows the model lightcurve from our best shape model, with a pole at ecliptic coordinates $(\lambda, \beta) = (52^{\circ}, -60^{\circ})$. Also shown, in red, is the lightcurve predicted by the model of \citen{Naidu2013}. Their nominal pole direction $(80^{\circ}, -50^{\circ})$ and shape model are incompatible with the lightcurve data. Plots of the Naidu et al. model and our model for the other lightcurves can be found in figures S2 through S9 in the Supplementary material.
}
\label{fig_lc_Feb24b_models}}
{\includegraphics[width=0.7\textwidth]{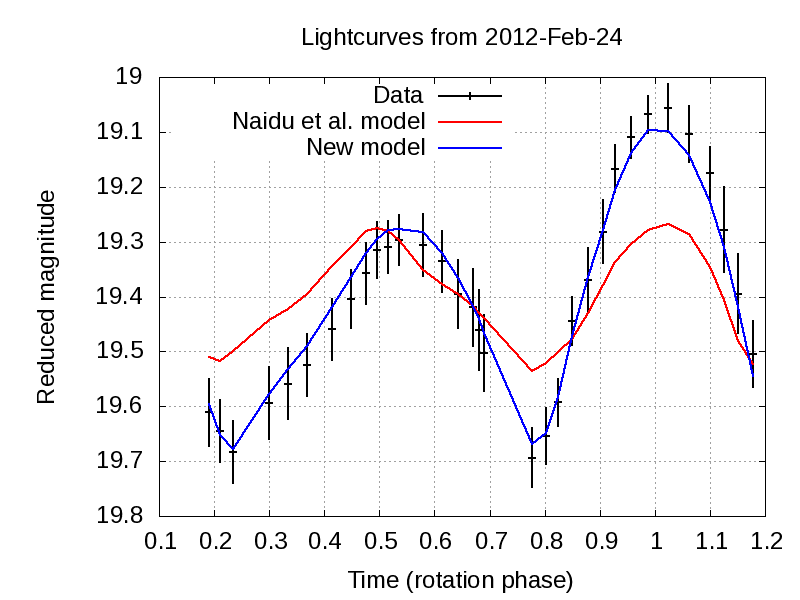}} %{lc_2012-Feb-24b_Kg_800x600.png}} % scale=0.4
\end{figure}

The lightcurves provide valuable information on ET70's shape and spin state. In particular, the times of their minima and maxima helped us refine ET70's rotation period, and their amplitudes constrained ET70's dimensions and pole position. There are two brightness maxima per rotation period. The maxima of several of the lightcurves are noticeably asymmetric -- that is, one of the maxima is clearly and consistently brighter than the other, as shown in \Fig{fig_lc_Feb24b_models}. This was a useful constraint for shape modeling, because some models could not reproduce the asymmetric maxima.

\clearpage
\section{Shape modeling \label{sec_shape}}

The general methodology for radar-based shape modeling was first described by \citen{Hudson1994} and is discussed in more detail by \citen{Magri2007, Magri2011}. In the shape modeling process, we use the code SHAPE, as described in these papers. SHAPE represents the asteroid's surface as a polyhedron with a series of triangular facets, finding the optimal value of each model parameter by calculating a noise-free simulated data set for each trial model and comparing it to the full set of actual data. Here, we briefly describe the key points of shape modeling for ET70.

We began with the published shape model of \citen{Naidu2013}, which has 2000~vertices and 3996~triangular facets. The average edge length of the triangles is about 100~meters. In the early stages of shape modeling, we used floating scale factors for the model's three principal axis lengths -- that is, we kept the general ``clenched fist'' shape of ET70 but allowed it to be stretched or compressed along each axis. This greatly reduced the required computational time, because it meant that SHAPE only had to find optimal values of the three axes' scale factors, instead of optimal displacements for each of the two thousand vertices. In the later stages of shape modeling, we did allow the individual vertices' displacements to vary.

The radar scattering properties of the asteroid were represented as a cosine law with respect to incidence angle: $\frac{d \sigma}{dA} = R (C+1)(\cos \theta)^{2C}$ \cite{Mitchell1996} where $\sigma$ is the radar cross section, $A$ is the surface area, $R$ and $C$ are fit parameters, and $\theta$ is the incidence angle. $\frac{d \sigma}{dA}$ is the radar cross section per unit area at an incidence angle $\theta$.

There were far more radar data points (hundreds of thousands of image pixels and spectral channels) than lightcurve data points (two hundred after decimation), so in order to ensure that each type of observation had significant leverage on the modeling results, we had to apply different weights to the various data sets. The final weights were set such that the delay-Doppler images contributed about half of total chi-squared, the CW spectra contributed about one third, and the lightcurves contributed about one sixth.

The Doppler bandwidth, $B$, of a continuous wave (CW) spectrum is given by $B = \frac{4 \pi D}{\lambda P} \cos \phi$ where $D$ is the projected breadth (diameter) of the asteroid, $\lambda$ is the wavelength of the radar, $P$ is the asteroid's rotation period, and $\phi$ is the sub-observer latitude. The key constraint from the CW spectra was on the sub-observer latitude at the times of the observations. Therefore our CW spectra, most of which are from a relatively narrow range of observation times, required ET70's pole to fall along a certain arc across the sky, but they did not allow the specific position on that arc to be determined.

We focused our efforts on examining pole positions near the nominal \citen{Naidu2013} retrograde solution at $(80^{\circ}, -50^{\circ})$. Testing showed that prograde solutions -- i.e., those near the conjugate pole $(260^{\circ}, +50^{\circ})$ -- are not compatible with the full set of radar and lightcurve data. All prograde models showed clear mismatches between the rotation phases of the model and the data. Most notably, the prograde model's rotation phase lags behind the data in later delay-Doppler images, but the model's rotation phase is ahead of the data in lightcurves from about the same observation times. Some of the delay-Doppler images were particularly helpful in constraining ET70's pole direction because they have two or more bright edges (see figures S12 through S34 in the Supplementary material), and a model with a misaligned pole would have those edges separated by the wrong number of delay cells.

\begin{figure}
\begin{tabular}{ccc}
\includegraphics[width=0.3\textwidth]{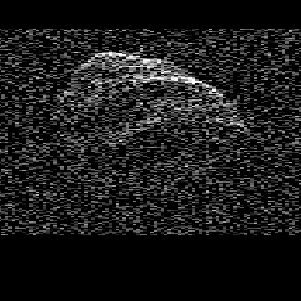} &
%{lon052_lat-60_sp965p800f_rp316p0f_Aug210f/write/pgm_ppm/obs_05_09_gamma2p0.png} % scale=0.4
\includegraphics[width=0.3\textwidth]{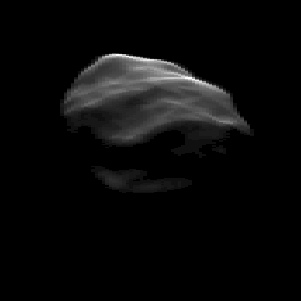} & %{lon052_lat-60_sp965p800f_rp316p0f_Aug210f/write/pgm_ppm/fit_05_09_gamma2p0.png}
\includegraphics[width=0.3\textwidth]{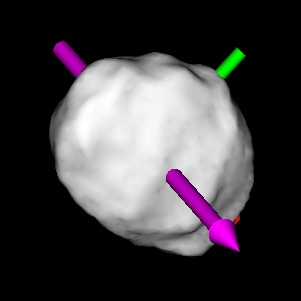} \\ %{lon052_lat-60_sp965p800f_rp316p0f_Aug210f/write/pgm_ppm/sky_05_09.png}
Delay-Doppler data & Model delay-Doppler image & Plane-of-sky view \\
\includegraphics[width=0.3\textwidth]{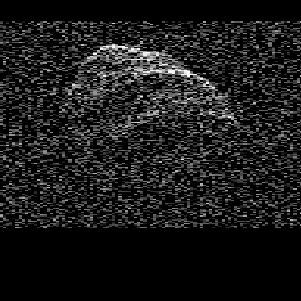} & %{lon040_lat-68_sp967p10f_rp300p0f_Aug000f/write/pgm_ppm/obs_05_09_gamma2p0.png}
\includegraphics[width=0.3\textwidth]{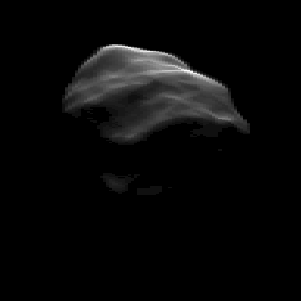} & %{lon040_lat-68_sp967p10f_rp300p0f_Aug000f/write/pgm_ppm/fit_05_09_gamma2p0.png}
\includegraphics[width=0.3\textwidth]{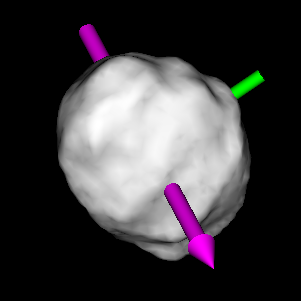} \\ %{lon040_lat-68_sp967p10f_rp300p0f_Aug000f/write/pgm_ppm/sky_05_09.png}
\end{tabular}
\caption{
An example of delay-Doppler frames for two models with different rotation periods and pole positions but nearly indistinguishable simulated images. The first row shows a model with a sidereal rotation period of 8.944~hours and a pole direction of $(52^{\circ}, -60^{\circ})$. The second row shows a model with a sidereal rotation period of 8.934~hours and a pole direction of $(40^{\circ}, -68^{\circ})$, which is $10^{\circ}$ from the first model's pole position.
The left column shows the data (same for both models, but with slightly different brightness scaling in the two frames). The center column shows the simulated (noise-free) delay-Doppler images derived from the shape models. In the delay-Doppler images, delay is plotted on the vertical axis (increasing from the top down), and Doppler frequency is plotted on the horizontal axis (increasing from left to right). The right column shows the simulated plane-of-sky views (as seen from Earth), with the sidereal spin vector (the model's shortest principal axis) shown as a magenta arrow and the long and intermediate principal axes shown as red and green shafts, respectively. In all plane-of-sky views, north is upward and east is leftward.
The radar images are stretched so that their spatial scale is the same as that of the plane-of-sky image (square frames with a side length of 3.5~kilometers), but the radar images' vertical axis, range, is perpendicular to the plane of the sky. Based on this work, the ET70 model's maximum extent is 2.9~km, and its minimum extent is 1.5~km. Note that the same region of the asteroid is closest to the observer for both models, so the model images are nearly identical. The main difference between these two models is a rotation in the plane of the sky, which does not affect delay-Doppler images. However, the models also differ by a slight rotation (about five~degrees) out of the plane.
}
\label{fig_DD_rot_set05}
\end{figure}

We also found that slight changes to the model's sidereal rotation period (of order $0.1\% \approx 0.009$~h~$\approx 30$~s), combined with a compensating change in ET70's pole position, of order $5^{\circ}$, produced simulated data sets that were practically indistinguishable from each other, and from the actual data (see \Fig{fig_DD_rot_set05}). In other words, with the available radar and lightcurve data, ET70's sidereal rotation period and pole direction are degenerate.

\begin{figure}
\begin{center}
\includegraphics[width=0.95\textwidth]{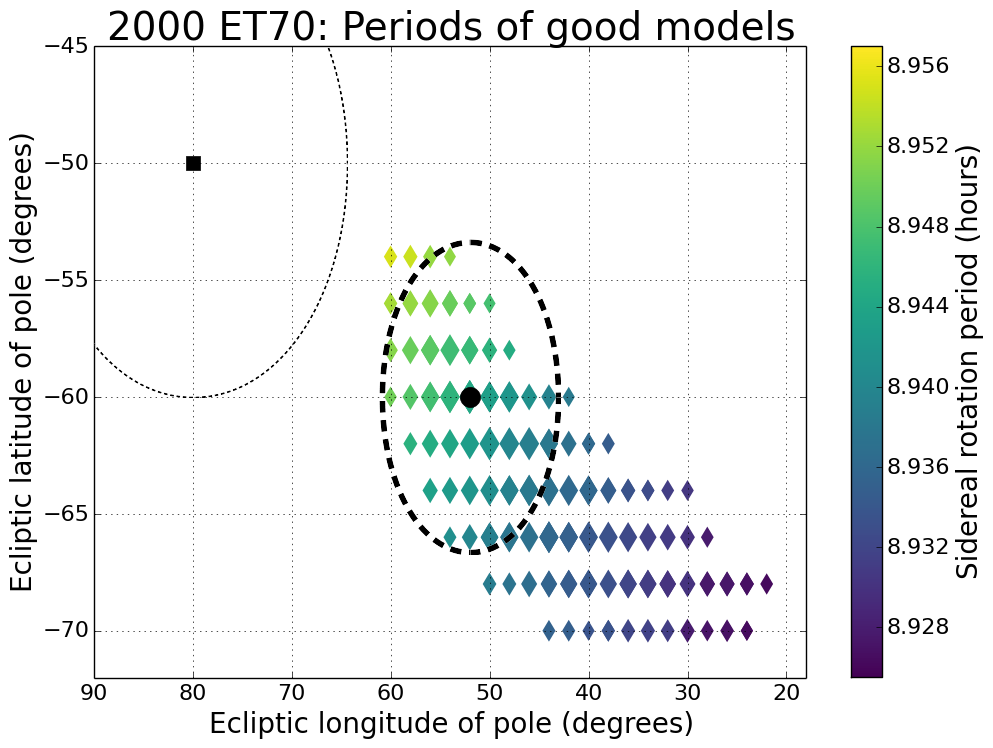} %{banana_plots_v21b/fig_banana_ellipses_period_v21b.png}
\end{center}
\caption{
The ``arc'' of good shape models' pole positions. Note that different pole positions require slightly different sidereal rotation periods.
The nominal pole solution $(80^{\circ}, -50^{\circ})$ of \citen{Naidu2013}, from using only the radar data, is indicated with a black square. The thin dashed ellipse shows their $10^{\circ}$ one-sigma uncertainty. This pole is ruled out by the lightcurves.
The colored diamonds show models which are compatible at the one-sigma level with the radar and lightcurve data (but not necessarily with the infrared spectra). Larger markers correspond to better shape models (that is, models with lower values of chi-squared).
Thermal modeling allows us to place additional constraints on the pole position, because some of the poles along the arc are ruled out by our infrared spectra (see \Sect{sec_thermal}). The points within the thick dashed ellipse are compatible at the one-sigma level with all available data (radar, lightcurve, and infrared). Pole positions on this ellipse are at the outer edges of acceptable thermal models. The black circle shows the pole solution $(52^{\circ}, -60^{\circ})$, which provides the best fits to the radar, lightcurve, and infrared data. Our preferred pole $(52^{\circ}, -60^{\circ})$ is $19^{\circ}$ from the pole of Naidu et al., which is about double their stated one-sigma uncertainty.
}
\label{fig_banana_period}
\end{figure}

In the final stages of shape modeling, we ran over 300 models, for which each model's pole position was held constant but the rotation period was allowed to vary. We found that the pole could fall within a region encompassing ecliptic longitudes $21^{\circ}$~to~$61^{\circ}$ and latitudes $-71^{\circ}$~to~$-53^{\circ}$, as shown in \Fig{fig_banana_period}. Different pole positions require slightly different sidereal rotation periods, ranging from 8.926~to~8.957~hours. The nominal shape model of this work provides slightly better fits to the radar data than that of \citen{Naidu2013}, but the main improvement is how it fits the lightcurves.
The nominal pole position of Naidu et al., $(80^{\circ}, -50^{\circ})$ with an uncertainty of $10^{\circ}$, can be ruled out because it is incompatible with the lightcurve data (see \Fig{fig_lc_Feb24b_models}, and also figures S2 through S9 in the Supplementary material). Our infrared observations provide additional constraints on the pole position, because thermal modeling shows that some poles and periods allowed by the radar and lightcurve data are ruled out by the infrared data (see \Sect{sec_thermal}).

\begin{figure}
\begin{center}
\includegraphics[width=\textwidth]{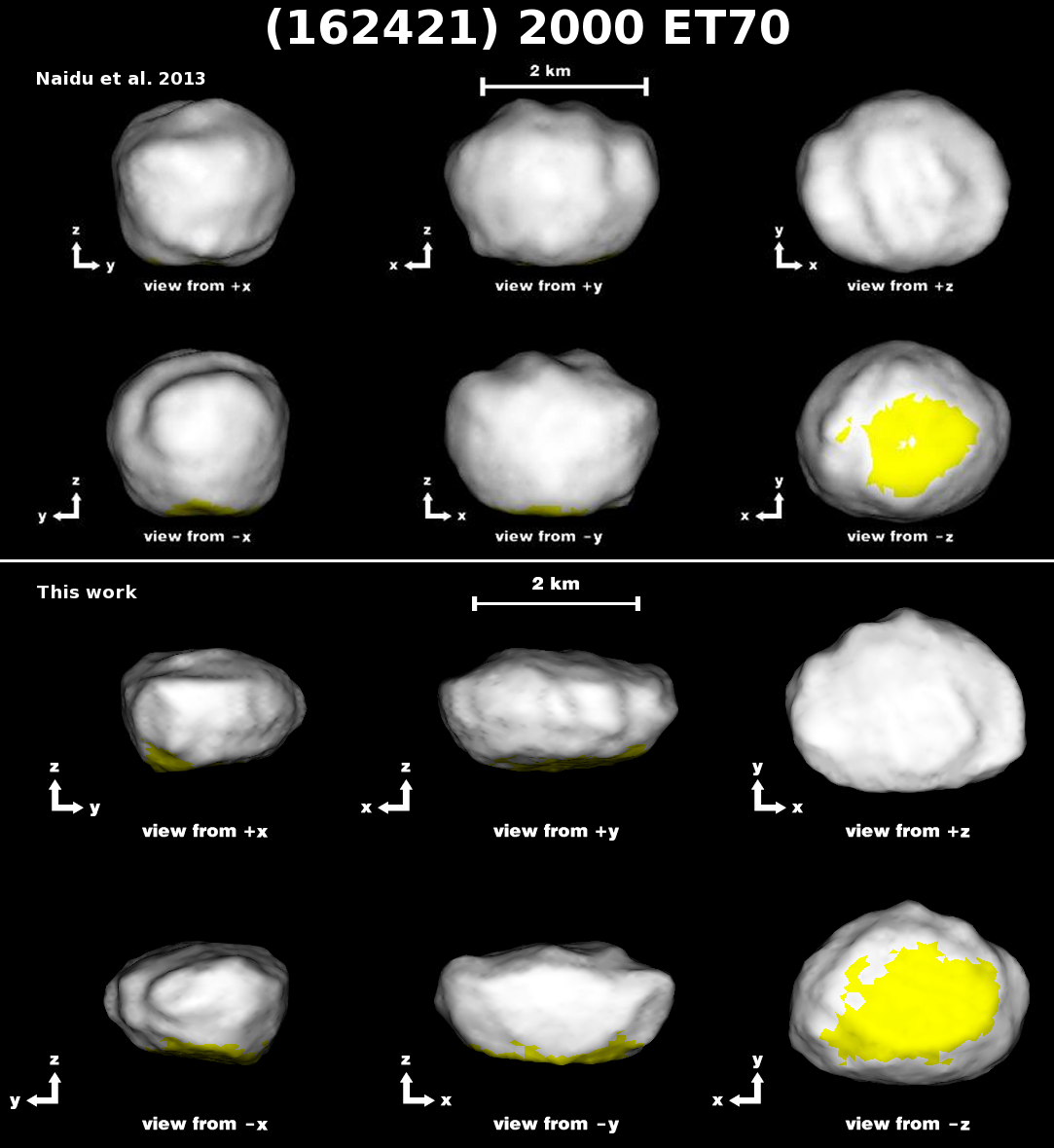} %{pa_collage_both_v17a.png} % scale=0.43
\end{center}
\caption{
The top frame shows views of the shape model of \citen{Naidu2013} from along its three principal axes (their Figure~7). The bottom frame shows the corresponding views of the best shape model from this work. The model rotates about the z-axis, and its maximum diameter is 2.9~kilometers. The facets marked in yellow, near the south pole, were seen at incidence angles greater than sixty~degrees in all delay-Doppler images (or not seen at all) and therefore are not well constrained.
}
\label{fig_shape_views}
\end{figure}

The final best-fit shape model has somewhat different dimensions from the Naidu et al. model, and there are also some small features that are slightly different (see \Fig{fig_shape_views}). The radar and lightcurve data favor a shape that is considerably shorter along its z-axis (rotation axis) than the radar-only model of \citen{Naidu2013}. This supports the suggestion of \citen{Rozitis2014} that shape models derived only from radar data may overestimate models' z-lengths, based on their analysis of 1620~Geographos (using infrared data to constrain the shape model). However, we note that the region near the southern pole was not seen clearly in the delay-Doppler images, and that ET70's z-length is not well constrained. The relative uncertainty in ET70's z-length is 25\%. The best-fit lengths along each axis vary with pole position (see figures S37 through S39 in the Supplementary material).

Figures~S37 through S39 in the Supplementary material show that models with a greater extent along the principal x-axis tend to be longer along the y-axis and shorter along the z-axis. The uncertainties for those three lengths are correlated, which affects the uncertainties in derived parameters \citep[e.g.][equation 3.13]{Bevington2003}. For the set of 72 good shape models which had pole positions that were compatible with thermal modeling results (that is, poles inside the thick dashed ellipse in \Fig{fig_banana_period}), covariances were calculated according to their definition, i.e. $\sigma_{xy} = \frac{1}{n-1} \sum\limits_{i=1}^n (x_i - \overline{x}) (y_i - \overline{y})$ \citep[][equation 3 and related discussion]{Hartlap2007}, and converted to correlation coefficients by $\rho_{xy} = \frac{\sigma_{xy}}{\sigma_{x} \sigma_{y}}$.
The correlation coefficient between x~and~y~is~+0.9, the correlation between x~and~z~is~-0.9, and the correlation between y~and~z~is~-0.9. These correlations affect the derived uncertainties in ET70's surface area, volume, moment of inertia ratios, and mean diameter; see \Tab{tab_shape_params}.

Our improved shape model's parameters and their uncertainties are given in \Tab{tab_shape_params}. The reported uncertainties in the model's parameters are conservative estimates based on combining changes in chi-squared with visual inspection of the models. As in \citen{Magri2007, Magri2011} and \citen{Nolan2013}, we ultimately had to make subjective decisions about what could be considered a good model. The preferred pole position is indicated in \Fig{fig_banana_period}. Principal axis views of the updated shape model are shown in \Fig{fig_shape_views}. The complete set of delay-Doppler frames and sums used in shape modeling are shown in figures S10 through S36 in the Supplementary material.

\begin{table}
\begin{tabular}{|lc|cr|cccr|}
\hline
\multicolumn{2}{|c|}{\multirow{3}{*}{Parameter}} & \multicolumn{2}{|c|}{\citen{Naidu2013}} & \multicolumn{4}{|c|}{This work} \\
\multicolumn{2}{|c|}{}                          & \multirow{2}{*}{Value} & \multirow{2}{*}{Unc.} & \multirow{2}{*}{Value} & \multicolumn{2}{c}{\multirow{2}{*}{Uncertainties}} & Rel. \\
\multicolumn{2}{|c|}{}                          & {}                     & {}                    & {}                     & \multicolumn{2}{c}{{}}                             & unc. \\\hline
\multirow{3}{*}{Extents along principal axes *} & $x$ & 2.61 km & 5\% & 2.90 km & -0.24 km & +0.34 km & 10\% \\
{}                                              & $y$ & 2.22 km & 5\% & 2.24 km & -0.23 km & +0.36 km & 13\% \\
{}                                              & $z$ & 2.04 km & 5\% & 1.50 km & -0.29 km & +0.47 km & 25\% \\\hline
Surface area * & $A$ & 16.7 km\texttwosuperior & 10\% & 15.3 km\texttwosuperior & \multicolumn{2}{c}{1.0 km\texttwosuperior} & 7\% \\\hline
Volume * & $V$ & 6.07 km\textthreesuperior & 15\% & 4.82 km\textthreesuperior & \multicolumn{2}{c}{0.50 km\textthreesuperior} & 10\% \\\hline
\multirow{2}{*}{Moment of inertia ratios *} & $I_x/I_z$ & 0.80 & 10\% & 0.53 & \multicolumn{2}{c}{0.11} & 20\% \\
{}                                          & $I_y/I_z$ & 0.96 & 10\% & 0.80 & \multicolumn{2}{c}{0.13} & 17\% \\\hline
Volumetric mean diameter *         & $D$ & 2.26 km & 5\% & 2.10 km & \multicolumn{2}{c}{0.07 km} & 3\% \\\hline
\multirow{3}{*}{DEEVE extents}    & $2a$ & 2.56 km & 5\% & 2.88 km & \multicolumn{2}{c}{0.29 km} & 10\% \\
{}                                & $2b$ & 2.19 km & 5\% & 2.18 km & \multicolumn{2}{c}{0.29 km} & 13\% \\
{}                                & $2c$ & 2.07 km & 5\% & 1.47 km & \multicolumn{2}{c}{0.37 km} & 25\% \\\hline
Radar albedo (2380 MHz, OC) & $\hat{\sigma}_{OC}$ & 0.063 & 0.017 & 0.062 & \multicolumn{2}{c}{0.018} & 30\%  \\\hline
Sidereal rotation period           & $P$ & 8.96 h & 0.01 h & 8.9444 h & -0.0081 h & +0.0100 h & 0.1\%  \\\hline
Pole ecliptic longitude & $\lambda$ & $80^{\circ}$ & \multirow{2}{*}{$10^{\circ}$} & $52^{\circ}$ & \multicolumn{2}{c}{$9^{\circ}$} & \\
Pole ecliptic latitude  &  $\beta$ & $-50^{\circ}$ & {}                           & $-60^{\circ}$ & \multicolumn{2}{c}{$6^{\circ}$} & \\
\hline
\end{tabular}
\caption{
Parameters and one-sigma uncertainties for the shape model of ET70, utilizing radar, lightcurve, and infrared data. The ranges of values given here are those for which both shape models and thermal models provide good fits to the data.
For many parameters, the uncertainties are asymmetric, so we include the values for both the negative and positive error bars. For instance, the model's extent along its principal x-axis has a best-fit value of 2.90~km, with a one-sigma range of 2.66~to~3.24~km.
Length uncertainties have contributions from two sources: there is uncertainty in the best-fit length (along each axis) for a fixed pole position, and the best-fit lengths vary with pole position (see figures S37 through S39 in the Supplementary material).
Relative uncertainties come from the mean of the positive and negative uncertainties.
The moment of inertia ratios were derived with the assumption that the model's density is homogeneous.
The dynamically equivalent equal volume ellipsoid (DEEVE) is the uniform-density ellipsoid that has the same volume and moment of inertia ratios as the shape model.\newline
* Our analysis showed that the uncertainties for the three lengths along the principal axes are correlated. The uncertainties in the model's surface area, volume, moment of inertia ratios, and mean diameter are therefore different from what they would be if the uncertainties in the three lengths were uncorrelated.
}
\label{tab_shape_params}
\end{table}

\clearpage
\section{Spectral observations \label{sec_obs_vnir}}

\subsection{Infrared observations from IRTF \label{sec_obs_ir}}

Infrared observations of an asteroid's thermal emission make it possible to determine the asteroid's albedo, surface roughness, and thermal inertia \citep[e.g.][]{Lebofsky1978, Harris2002}. As part of our ongoing program to characterize near-Earth asteroids with both radar and infrared observations, we observed ET70 on three nights in February~2012 from NASA's InfraRed Telescope Facility (IRTF) on Mauna Kea, Hawai'i (see \Tab{tab_obs_ir}). All of our infrared observations used the SpeX instrument \cite{Rayner2003} in two modes, PRISM and LXD, giving coverage from 0.8~to~4.1~microns.

\begin{table}
\begin{tabular}{|lccrr|cccc|}
\hline
\multirow{2}{*}{Date} & \multirow{2}{*}{UT times} & Instrument & Exposure & \multirow{2}{*}{Exposures} & Standard & $r$ & $\Delta$ & \multirow{2}{*}{$\alpha$} \\
{} & {} & mode       & time (s) & {} & stars & (au) & (au) & {} \\\hline
                            &              &           &    &    & SAO 65083, &    &    &    \\
\multirow{2}{*}{2012-02-11} & 14:13--14:53 & LXD       & 15 & 72 & SAO 82194, & \multirow{2}{*}{1.008} & \multirow{2}{*}{0.071} & \multirow{2}{*}{$71^{\circ}$} \\
{}                          & 15:00--15:06 & PRISM     & 30 &  8 & SAO 83619, & {} & {} & {} \\
                            &              &           &    &    & SAO 98710  &    &    &    \\\hline
\multirow{6}{*}{2012-02-18} & 13:39--13:45 & PRISM (A) & 30 &  8 &            & \multirow{6}{*}{1.019} & \multirow{6}{*}{0.046} & \multirow{6}{*}{$47^{\circ}$} \\
{}                          & 13:51--14:01 & LXD (A)   & 15 & 18 &            & {} & {} & {} \\
{}                          & 14:05--14:29 & LXD (B)   & 15 & 40 & SAO 98710, & {} & {} & {} \\
{}                          & 14:29--14:47 & LXD (C)   & 15 & 32 & SAO 120107 & {} & {} & {} \\
{}                          & 14:54--15:00 & PRISM (B) & 30 &  8 &            & {} & {} & {} \\
{}                          & 15:11--15:33 & LXD (D)   & 15 & 40 &            & {} & {} & {} \\\hline
\multirow{3}{*}{2012-02-21} & 13:12--13:25 & LXD (A)   & 15 & 24 & SAO 98710, & \multirow{3}{*}{1.024} & \multirow{3}{*}{0.047} & \multirow{3}{*}{$41^{\circ}$} \\
{}                          & 13:28--13:33 & PRISM (A) & 30 &  8 & SAO 120107, & {} & {} & {} \\
{}                          & 13:43--13:49 & LXD (B)   & 15 & 12 & SAO 180396 & {} & {} & {} \\\hline
\end{tabular}
\caption{
Infrared observations of 2000~ET70 that were used in thermal modeling. All of these observations used the SpeX instrument on NASA's InfraRed Telescope Facility (IRTF).
The two modes of observations, PRISM and LXD, cover different ranges of wavelengths. PRISM covers 0.8~to~2.5~$\mu$m, and LXD (cross-dispersed) covers 2.2~to~4.1~$\mu$m.
The observations are grouped in terms of which sum each set of exposures went into, which is why there may be consecutive sets with the same mode. For instance, the LXD observations from 13:51~to~14:47 on February~18 were divided into three sums (called A, B, and C).
$r$ is the distance from the asteroid to the Sun; $\Delta$ is the distance from the asteroid to Earth; $\alpha$ is the solar phase angle.
The model's orientations at the times of the LXD observations are shown in \Fig{fig_tsky_colorized}.
}
\label{tab_obs_ir}
\end{table}

Our observations were taken using the standard method of A-B pairs, where the telescope moved fifteen~arcseconds along the slit between exposures, so that the target alternated between two positions within the slit. This allowed for a clean sky subtraction while still integrating on the asteroid. In addition to observing the asteroid, we observed solar-type comparison stars within five~degrees of the target to match airmass as accurately as possible, as well as known solar analog stars (normally not as close to the target).
We used the solar analog star's spectrum to modify the colors of the solar-type comparison star, in order to make the comparison star's spectrum closer to the solar spectrum. We processed our SpeX data using the Spextool software \cite{Cushing2004}, along with Bus's method of correcting for telluric water vapor in PRISM spectra \citep[described in][]{Rivkin2004} with some minor modifications. Similarly, we correct for telluric features in the LXD spectra as described by \citen{Volquardsen2007}. Our procedures for infrared observation and data reduction are discussed in further detail in \citen{Howell2017}.
Example spectra are shown in \Sect{sec_thermal} (and also in the Supplementary material). Note that our infrared observations are obtained as relative reflectance as this is more robust to observing conditions than absolute photometry.

\subsection{Spectral classification \label{sec_spec_class}}

\begin{figure}
\floatbox[{\capbeside\thisfloatsetup{capbesideposition={right,center}}}]{figure}[\FBwidth]
{\caption{
Reflectance spectrum of 2000~ET70, combined from Palomar DBSP and IRTF SpeX PRISM observations. The green X's use the data points from the ECAS spectrum obtained by \citen{Whiteley2004} for wavelengths of less than 1.0~microns.
}
\label{fig_taxonomic_class}}
{\includegraphics[width=0.7\textwidth]{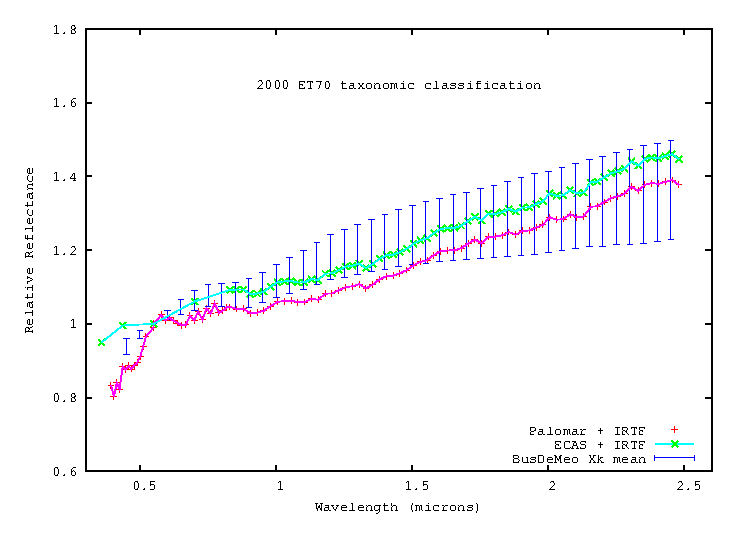}} %{ET70_taxonomy_Ellen_July11_mod.png} % scale=0.5
\end{figure}

\citen{Whiteley2001} classified ET70 as an X-type asteroid in the \citen{Tholen1984} taxonomy, based on ECAS photometry (0.3~to~1.0~microns, shown in \Fig{fig_taxonomic_class}) that is flat or slightly red with respect to the Sun. The Tholen X-types are separated by albedo into the E, M, and P classes, and the low albedo that we find for ET70 (in \Sect{sec_conc}) is only consistent with the P~class.

ET70 was observed by two of the authors (Hicks and Lawrence) at the Palomar 5.1-meter telescope on February~2,~2012, using the Double Spectrograph (DBSP). The blue and red portions of the spectrum were measured simultaneously, giving coverage from 0.4~to~1.0~microns, and the two portions were scaled to match in the region of overlap.

Using more extended spectral coverage (0.4~to~2.5~microns), with the thermal contribution removed, we have classified ET70 as Xk, in the Bus-DeMeo system \cite{DeMeo2009} (see \Fig{fig_taxonomic_class}). Depending on whether we scale the spectra to the Palomar spectrum, or the photometry of \citen{Whiteley2004}, the relative reflectance in the near-infrared region can vary by about 8\%. Measurements of the 0.8-2.5~micron region on the three different IRTF nights in February~2012 are consistent with each other to $\pm0.015$.

The photometry by \citen{Whiteley2004} and the visible spectrum from Palomar diverge at short wavelengths, but they were obtained twelve years apart, and the relative orientation of the object is not known. The range of values for other Xk asteroids falls between the two. If the spectra are normalized (matched) at 1.0-1.5~microns, the visible spectra are more consistent with each other, outside of the value at 0.55~microns. We do not have any reason at this time to consider one or the other to be more reliable. Inhomogeneity in surface composition could result in both variable thermal properties and in variable spectral colors, and additional future observations could explore this possibility.

\section{Thermal modeling \label{sec_thermal}}

In order to make the connection between the asteroid's thermal properties and the observed infrared spectra, we use a thermal model based on the derived shape model. The thermal model specifies how to calculate the temperatures across the asteroid's surface, based on the actual solar illumination and the asteroid's properties \cite{Morrison1973, Spencer1989, Lagerros1996, Harris2002}. The modeled thermal spectra of the regions visible to the observer are then summed to find the disk-integrated spectra that would be observed, which can be compared to the measured spectra. We use our detailed shape model to provide a more accurate representation of the thermal emission.

For our thermophysical modeling code, SHERMAN, we specify the asteroid's physical properties, including its size, shape, and rotation state, and fit for its optical scattering law, thermal inertia, and surface roughness; see \citen{Magri2017} and \citen{Howell2017} for more details. The model's infrared emissivity is assumed to be 0.9, a typical value for silicate minerals \cite{Brown1982, Spencer1989, Mueller2007}.

The asteroid's shape is represented as a polyhedron with triangular facets (the output from SHAPE). We calculate the asteroid's temperature distribution based on how the various parts of the asteroid's surface are illuminated by the Sun, solving the heat equation at closely spaced time steps to represent vertical heat transport (conduction and radiation) into and out of the subsurface layers beneath each facet (horizontal heat transport is ignored). We begin calculating the asteroid's thermal state several rotations before the observation times, in order to ensure that the model's thermal state has stabilized by the observation times.
Using the asteroid's actual shape instead of a sphere lets the modeling account for large-scale roughness, such as the ridges and valleys near the ET70 model's north pole. However, these valleys (the model's largest concavities) were mostly out of the sunlight during the times of our infrared observations, so they did not make a large contribution to the observed disk-integrated spectra.

Surface roughness on scales smaller than the facet size (i.e., ten-meter scales and smaller) is modeled by having some fraction of the asteroid's surface covered by a set of spherical-section craters. This parameterization of roughness follows the method of \citen{Lagerros1998}, who found that such craters give similar results to more complicated representations of surface roughness (for disk-integrated spectra), while requiring far less computational effort. SHERMAN allows sunlight to scatter multiple times within the crater, as per the assumed Hapke law. It then uses the absorbed fluxes to calculate the surface temperatures in the crater, following \citen{Lagerros1998}, with mutual heating (infrared emission and absorption) taken into account but assuming zero thermal inertia. The temperatures within the crater are then corrected for finite thermal inertia.

Modifying the notation of \citen{Lagerros1998}, we denote the crater coverage fraction as $f_c$ and the RMS slope as $\rho$. (Note that $\rho$ is a dimensionless slope; $\rho = \tan \theta_{rms}$.) We use $\gamma$ to denote the craters' opening angle. An opening angle of $\gamma = 180^{\circ}$ would indicate craters that are hemispheres, and $\gamma < 180^{\circ}$ indicates shallower craters. $\rho$, $f_c$, and $\gamma$ are related by
$\rho^2 = f_c \frac{\ln (1 - 2 S) - 2 S (S - 1)}{4 S (S - 1)}$
where $S = \frac{1 - \cos (\gamma/2)}{2} = \sin^2 (\gamma/4)$ is the ratio of the crater's depth to the diameter of its defining sphere.

For the thermal models of ET70, we used a crater opening angle of $\gamma = 150^{\circ}$, similar to angles used for the `default roughness' and `high roughness' cases of \citen{Mueller2007}. We varied the crater fraction $f_c$ but kept the opening angle fixed, because \citen{Lagerros1998} and \citen{Emery1998} found that combinations of $\gamma$ and $f_c$ that equate to the same average roughness produce nearly indistinguishable model spectra.
Because we used ET70's actual shape and explicitly allowed the model's sub-facet surface roughness to vary, we did not need to separately incorporate a beaming parameter ($\eta$), which has been used by many previous studies to incorporate effects of anisotropic scattering, non-spherical shape, thermal inertia, surface roughness, and other inhomogeneous surface properties \citep[e.g.][]{Lebofsky1986}.

Optical scattering was represented with a Hapke law \cite{Hapke1984}. In order to reduce the dimensionality of the parameter space, we used two free parameters, visual albedo ($p_V$) and phase slope parameter ($G$). We converted $p_V$ and $G$ to Hapke parameters using the formulas from \citen{Verbiscer1995}. A lower value of $G$ corresponds to a stronger opposition surge, but since all observations of ET70 were taken at phase angles greater than 40~degrees, the data do not constrain ET70's opposition surge. Testing showed that changing the value of $G$ and the corresponding Hapke parameters did not significantly change the thermal modeling results, so we allowed the single-scattering albedo ($w$) to vary but kept the other Hapke parameters fixed at the values derived from $G = 0.17$, which is the average value for Xk-type asteroids \cite{Warner2009}.

\subsection{Reflectance spectrum \label{sec_thermal_refl}}

Careful analysis of our infrared data allowed us to generate the asteroid's reflectance spectrum. At PRISM wavelengths (0.8~to~2.5~microns), most of the observed flux is from reflected sunlight, rather than thermal emission. However, the thermal contribution becomes significant (larger than the data's error bars) above wavelengths of about 2.2~microns. Therefore we had to remove the thermal component from the observed PRISM spectra in order to produce the reflectance spectrum that was used for SHERMAN. Our first estimate of the thermal contribution at the various PRISM wavelengths was based on the results from some early thermal models. However, this resulted in thermal models in which the model spectra consistently had higher values than what we observed at wavelengths near 2.4~microns, so further corrections were needed.

We assumed that the reflectance spectrum is flat above a certain wavelength. This cutoff wavelength was determined iteratively, by testing thermal models with different versions of the reflectance spectrum to see which cutoff wavelength would yield model PRISM spectra (with reflected and thermal contributions) that best matched the data. Other representations of the reflectance spectrum are possible -- for instance, assuming a constant slope out to some point -- but these would require additional parameters. Since less than 5\% of the power in the solar spectrum comes from wavelengths above 2.2~microns, the details of the parameterization are not critical, so we opted for simplicity.
We found that a cutoff wavelength of 2.25~microns is optimal. For comparison, the initial version of the reflectance spectrum had its cutoff at 2.48~microns. Different versions of the reflectance spectrum are shown in Supplementary online Figure~S40. Note that, once a good reflectance spectrum is used, the PRISM spectra are not very sensitive to changes in the thermal parameters, so most of the leverage for the thermal models comes from the LXD spectra.

\subsection{Parameter search \label{sec_thermal_par}}

For thermal modeling, our three primary free parameters were the models' values of Hapke single-scattering albedo ($w$), crater coverage fraction ($f_c$), and thermal inertia ($\Gamma$). One can consider chi-squared to be a function of the thermal parameters; thermal modeling is effectively a minimization of this function $\chi^2(w, f_c, \Gamma)$. Given that ET70's pole position was uncertain, we also had to vary the pole position. This effectively added two more dimensions to search: the longitude and latitude within that arc of allowed pole directions (see \Fig{fig_banana_period}).
We ran thermal models for seven different pole positions. For each pole, we used the lengths and the sidereal rotation period that provided the best fit (from shape modeling) to that pole position. For one pole, we also tested thermal models for a shape with a greater length along its z-axis to see whether that would give better fits; it did not make a significant difference.

\begin{figure}
\begin{center}
\includegraphics[width=0.70\textwidth]{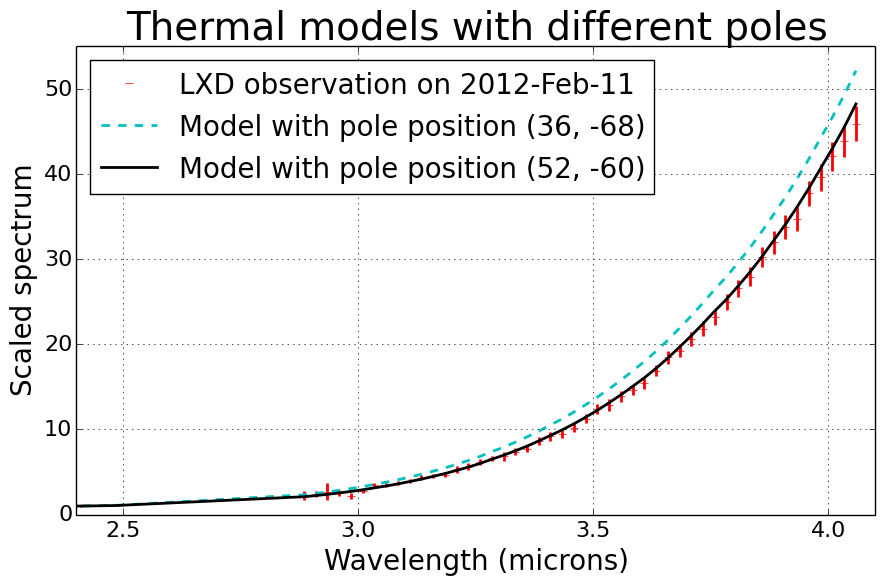} %{L11_models_different_poles_plot_v17a.png} % scale=0.5
\includegraphics[width=0.25\textwidth]{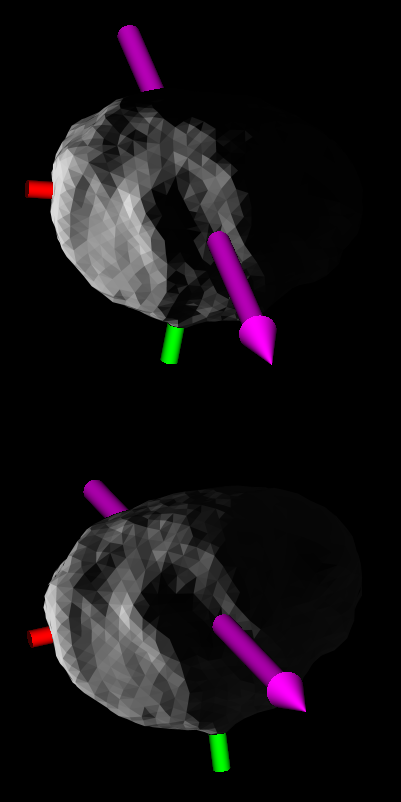} %{L11_models_different_poles_PoS_v17a.png} % scale=0.3
\end{center}
\caption{
Thermal models of 2000~ET70 with different pole positions. Because the models' pole positions differ, the latitude of the subsolar point changes, and different facets near the north pole are in shadow. These changes affect the disk-integrated spectra seen by the observer. The data points and their uncertainties are plotted in red. The first model, plotted in cyan, has a pole position of $(36^{\circ}, -68^{\circ})$, a sidereal rotation period of 8.932~hours, a single-scattering albedo of 0.13, a crater coverage fraction of 10\%, and a thermal inertia of 15~J~m\textsuperscript{-2}~K\textsuperscript{-1}~s\textsuperscript{-1/2}. The second model, plotted in black, has the nominal pole position $(52^{\circ}, -60^{\circ})$, a sidereal rotation period of 8.944~hours, a Hapke single-scattering albedo of 0.10, a crater coverage fraction of 10\%, and a thermal inertia of 65~J~m\textsuperscript{-2}~K\textsuperscript{-1}~s\textsuperscript{-1/2}.
Both pole positions are within the region shown in \Fig{fig_banana_period} that is compatible with the radar and lightcurve data. The first model is the best thermal model (lowest overall chi-squared) for its pole position, but it yields a poor fit, and therefore its pole can be ruled out because of thermal modeling.
The frames on the right show simulated plane-of-sky views of the two models (as seen from Earth), respectively, where each facet's brightness is proportional to its thermal emission. The magenta arrows are the models' rotation axes (their shortest principal axis). The red and green shafts are the long and intermediate principal axes, respectively. In the plane-of-sky views, north is upward and east is leftward.
}
\label{fig_therm_poles}
\end{figure}

For each tested pole position, we typically ran several dozen thermal models, to find the optimal values of $w$, $f_c$, and $\Gamma$, along with their allowed ranges. Ultimately, thermal modeling provided useful constraints on ET70's pole direction, because some pole directions that are allowed by the radar and lightcurve data had no thermal models that gave an acceptable fit to our IRTF spectra (see \Fig{fig_therm_poles}). The radar and lightcurve data allow for pole positions along an arc that is about $23^{\circ}$ long; thermal modeling showed that only about $12^{\circ}$ of that arc are compatible with the infrared spectra (see \Fig{fig_banana_period}).

The region of pole directions that is compatible with the thermal models is centered on a pole position of $(52^{\circ}, -60^{\circ})$. The corresponding sidereal rotation period for that nominal pole is 8.944~hours. Assuming a surface with homogeneous properties, some of the best thermal models had a crater fraction of 0.0 -- that is, no sub-facet surface roughness. This seems physically unlikely, although \citen{Naidu2013} noted that ET70's radar circular polarization ratio is lower than that of most NEAs, indicating a relatively smooth surface at scales of about 10~cm. However, this could be just a coincidence, because the roughness that affects thermal emission could be at spatial scales anywhere between the diurnal thermal skin depth (millimeters) and the size of the facets (tens of meters) \cite{Lagerros1998, Mueller2007}.

No single homogeneous thermal model could provide an acceptable fit to the infrared spectra from all three nights simultaneously; models that fit well for one night were poor for the other nights. For February~18, the situation broke down even further in that fits to data from earlier in the night (LXD sets A, B, and C) required models that were different from the models that fit data from later in the night (LXD set D).

\begin{figure}
\floatbox[{\capbeside\thisfloatsetup{capbesideposition={right,center}}}]{figure}[\FBwidth]
{\caption{
The best-fit values of crater fraction and thermal inertia for each subset of the data, for $w = 0.10$ and $G = 0.17$, with the pole at the nominal $(52^{\circ}, -60^{\circ})$. Models whose parameters fall along a single subset's line are nearly indistinguishable for that particular subset, illustrating that a family of solutions exists for each thermal spectrum.
}
\label{fig_therm_params_lines}}
{\includegraphics[width=0.4\textwidth]{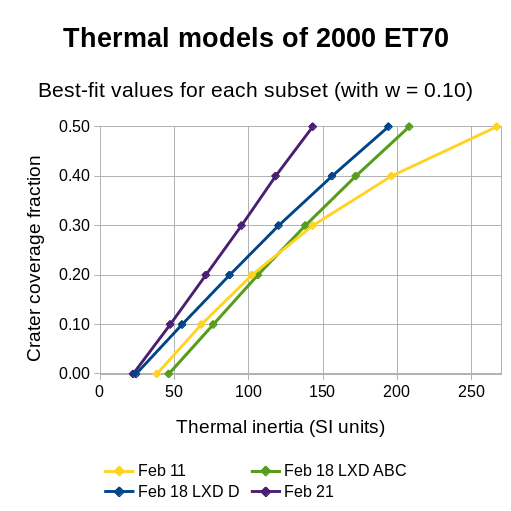}} %{Best_models_v17.png} % scale=0.6
\end{figure}

Focusing on four individual subsets of the infrared data (Feb-11, Feb-18 LXD~ABC, Feb-18 LXD~D, and Feb-21), we found that within a given subset there is a range of thermal parameters that fit the spectrum from that subset reasonably well, indicating that we do not have a single ``best fit'' in the classical sense but rather a family of solutions in each case. This is illustrated in \Fig{fig_therm_params_lines} for a single value of the single-scattering albedo $w$: once $w$ is set, there are curves of crater fraction and thermal inertia that result in nearly indistinguishable model spectra for each subset. To either side of a curve, the chi-squared value falls off rapidly, indicating that the models along a given curve are well constrained, even if the particular choice of curve is not. However, the more important point is that the families of solutions do not all overlap at any point, indicating that the thermal parameters are changing across the surface of the asteroid.

\begin{figure}
\begin{center}
\includegraphics[height=0.80\textheight]{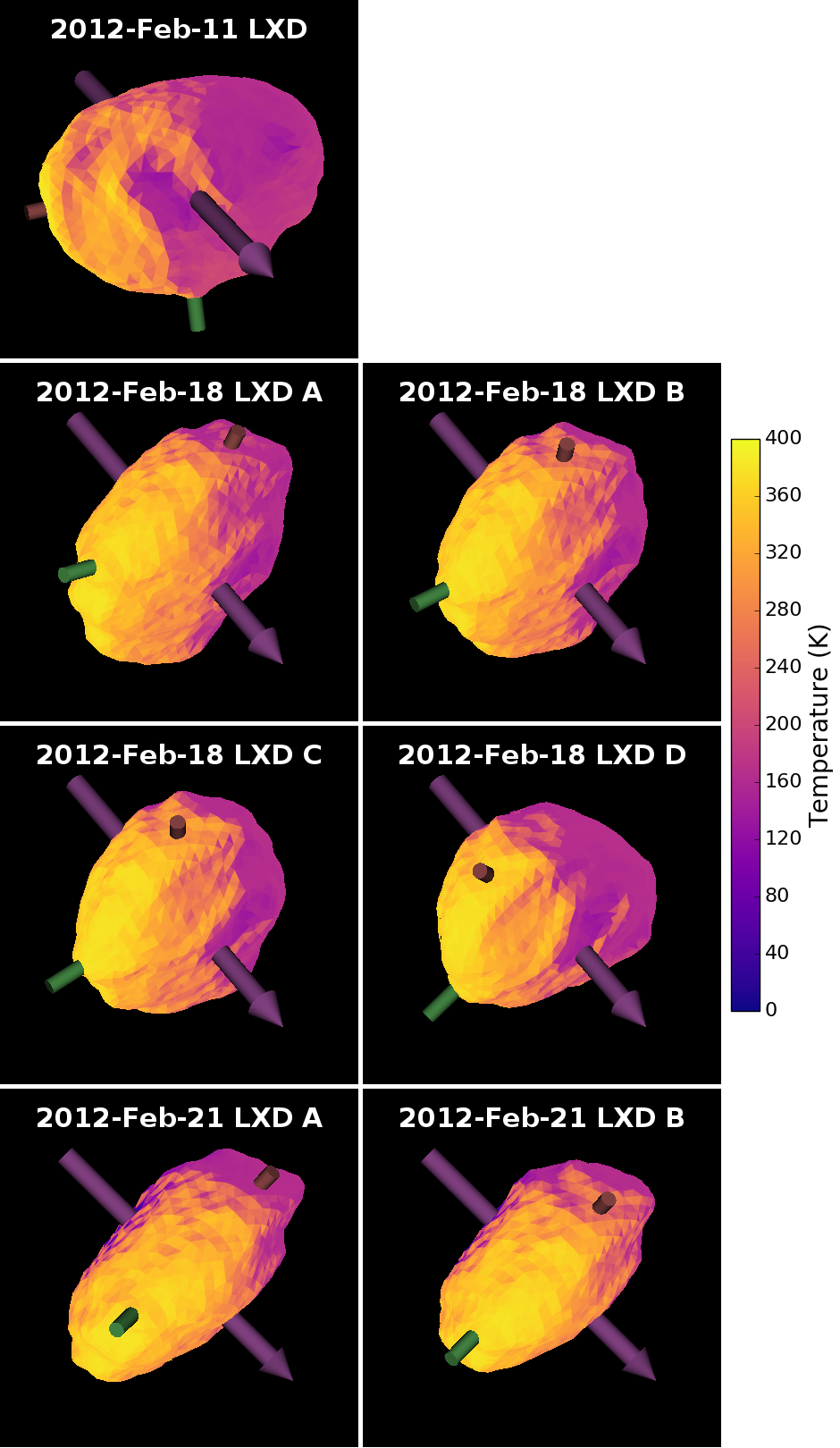} %{tsky_colorized/tsky_plasma_collage.png}
\end{center}
\caption{
Plane-of-sky views (as seen from Earth) for the seven individual LXD time periods using the final shape model and spin state. North is upward and east is leftward. Colors indicate surface temperatures based on a thermal model with $w = 0.108$, $f_c~=~0.2$, and $\Gamma~=~70$~J~m\textsuperscript{-2}~K\textsuperscript{-1}~s\textsuperscript{-1/2}, an arbitrary choice with parameter values that are in the middle of the allowed regions for all dates (see \Fig{fig_therm_Feb11} through \Fig{fig_therm_Feb21}). Note that most of the observed thermal emission at any given time comes from a small region on the surface, so observations at different times are dominated by the thermal properties of that region.
}
\label{fig_tsky_colorized}
\end{figure}

\begin{figure}
\begin{center}
\includegraphics[height=0.35\textheight]{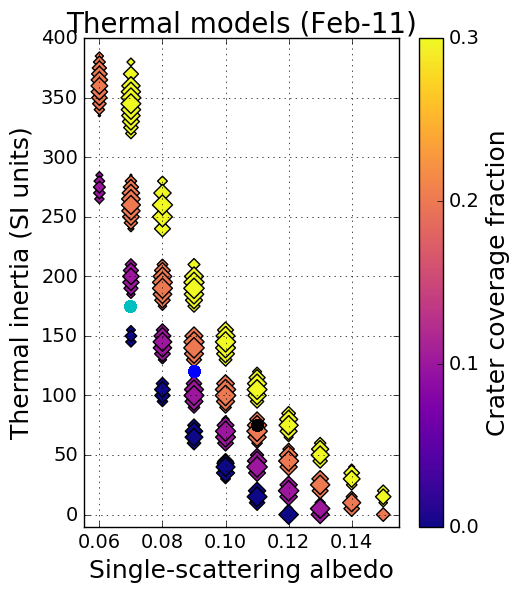} \includegraphics[height=0.35\textheight]{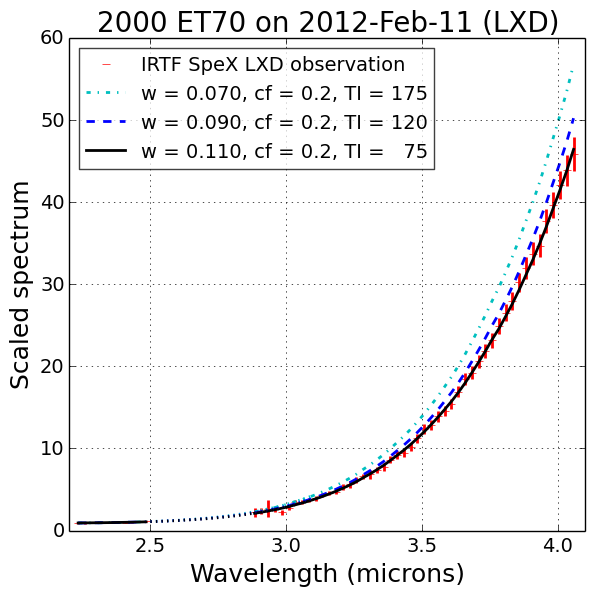} %{thermal_model3_plots_b15/v7b_plot_v7_2012-Feb-11a_LXD.png} % scale=0.6; width=0.75\textwidth
\end{center}
\caption{
Thermal models of ET70 for February~11, 2012.\newline
(Left) Illustration of the range of acceptable thermal models for this subset of data. Each point corresponds to a different model. The markers' colors indicate their values of crater coverage fraction. Larger markers correspond to better thermal models (i.e., those with lower chi-squared for the February~11 spectra). For a given crater fraction, there is a ``curve'' of models in which thermal inertia and albedo are traded off to be compatible with the infrared spectra.
Note that models are only shown for crater fractions of 0.3 or less. Models with greater values of crater fraction may be compatible with certain subsets of spectra, but they are much worse overall. That is, for higher crater fractions, it takes a much wider range of other parameter values to match the observations. For similar reasons, these plots only show thermal inertia values up to 400~J~m\textsuperscript{-2}~K\textsuperscript{-1}~s\textsuperscript{-1/2} and single-scattering albedo values from 0.06~to~0.15. The three dots (cyan, blue, and black) correspond to the models plotted in the right frame.\newline
(Right) Comparison of infrared observations to model LXD spectra. The model shown with the solid black curve fits well for Feb-11 and for Feb-18 LXD~ABC, but not for Feb-18 LXD~D or for Feb-21.
}
\label{fig_therm_Feb11}
\end{figure}

\begin{figure}
\begin{center}
\includegraphics[height=0.35\textheight]{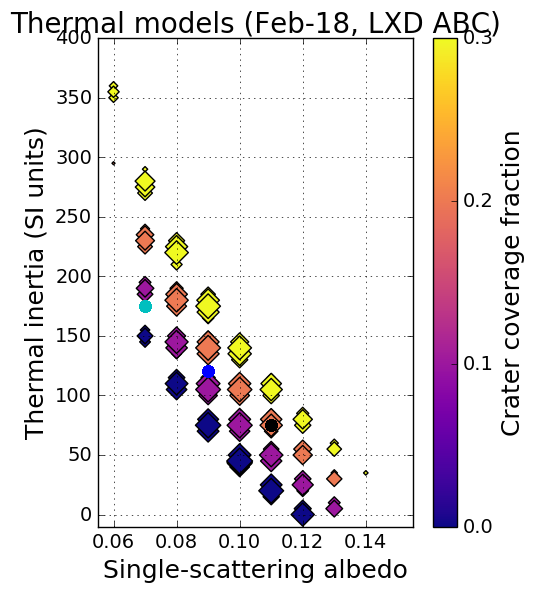} %{chi_squared_scatter_v5c/markers/w_ti_cf__rcs_Feb-18l_LXD_ABC_plasma.png}
\includegraphics[height=0.35\textheight]{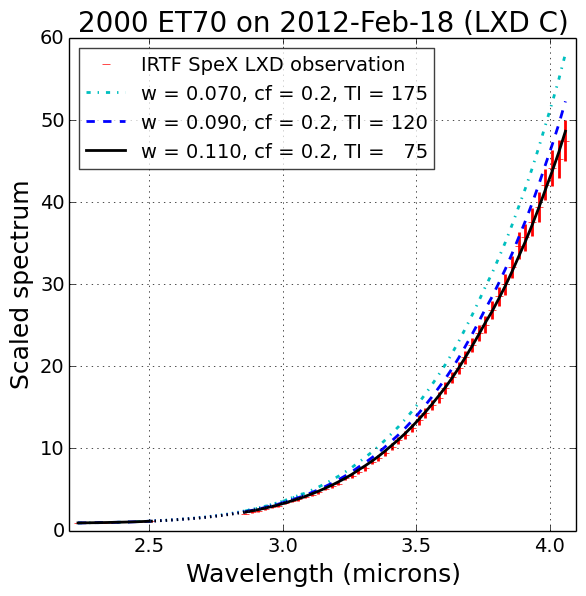} %{thermal_model3_plots_b15/v7b_plot_v7_2012-Feb-18d_LXD_C.png} % scale=0.6; width=0.75\textwidth
\end{center}
\caption{
Same as \Fig{fig_therm_Feb11}, but for the first three LXD observations from February~18, 2012. On the right, the model shown with the solid black curve fits well for Feb-11 and for Feb-18 LXD~ABC, but not for Feb-18 LXD~D or for Feb-21.
}
\label{fig_therm_Feb18_Labc}
\end{figure}

\begin{figure}
\begin{center}
\includegraphics[height=0.35\textheight]{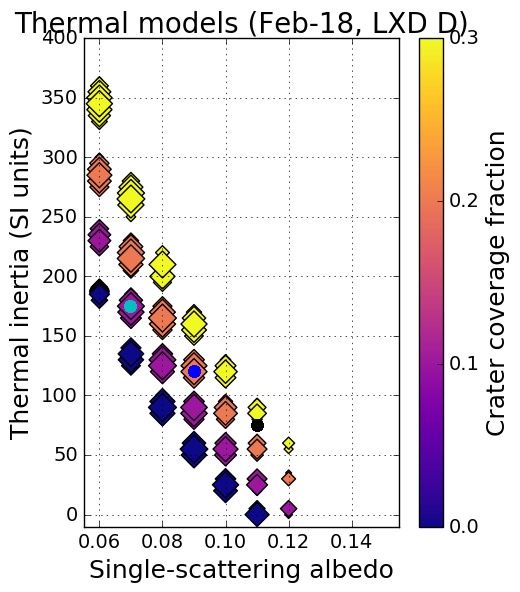} %{chi_squared_scatter_v5c/markers/w_ti_cf__rcs_Feb-18f_LXD_D_plasma.png}
\includegraphics[height=0.35\textheight]{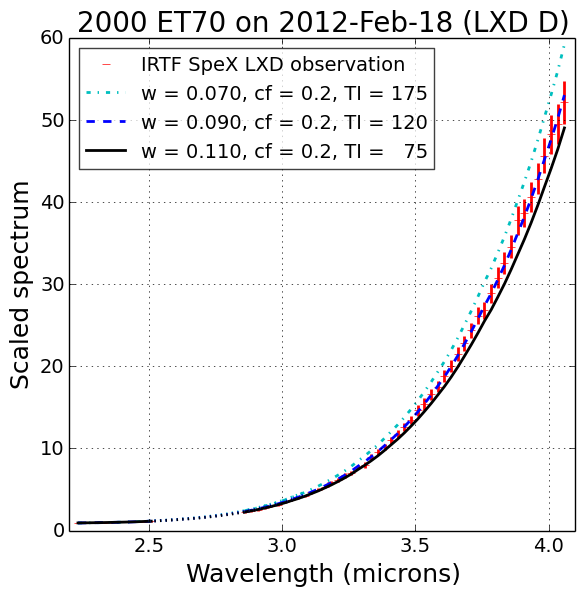} %{thermal_model3_plots_b15/v7b_plot_v7_2012-Feb-18f_LXD_D.png}
\end{center}
\caption{
Same as \Fig{fig_therm_Feb11}, but for the last LXD observation from February~18, 2012. On the right, the model shown with the dashed blue curve only fits well for Feb-18 LXD~D.
}
\label{fig_therm_Feb18_Ld}
\end{figure}

\begin{figure}
\begin{center}
\includegraphics[height=0.35\textheight]{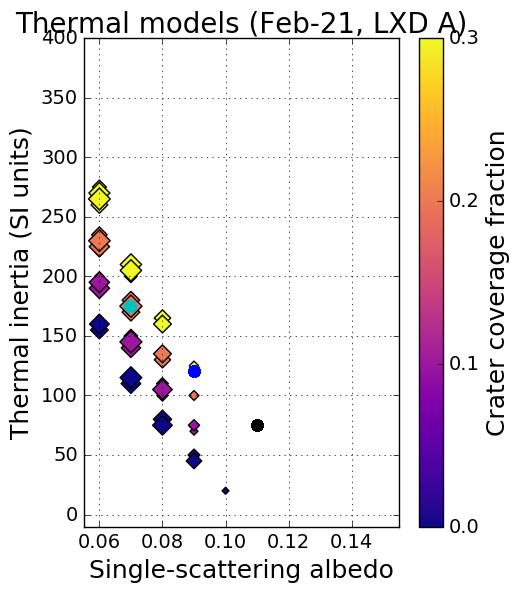} %{chi_squared_scatter_v5c/markers/w_ti_cf__rcs_Feb-21a_LXD_A_plasma.png}
\includegraphics[height=0.35\textheight]{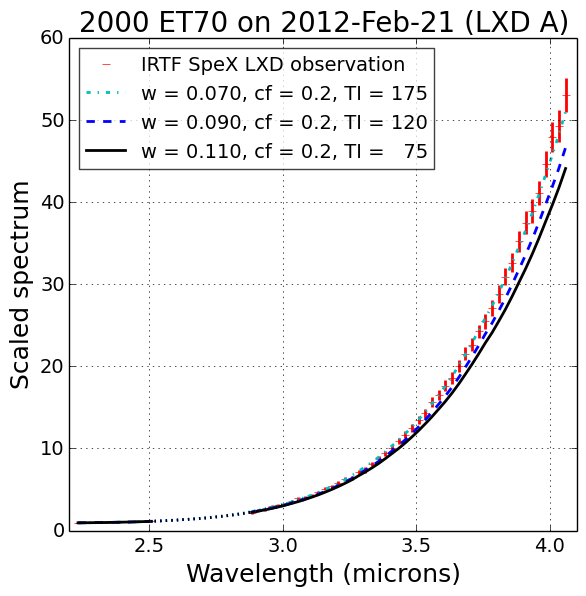} %{thermal_model3_plots_b15/v7b_plot_v7_2012-Feb-21a_LXD_A.png}
\end{center}
\caption{
Same as \Fig{fig_therm_Feb11}, but for the first LXD observation from February~21, 2012. On the right, the model shown with the dot-dashed cyan curve only fits well for Feb-21.
}
\label{fig_therm_Feb21}
\end{figure}

This is made clearer in \Fig{fig_tsky_colorized} through \Fig{fig_therm_Feb21}. \Fig{fig_tsky_colorized} shows the plane-of-sky views of the asteroid shape model during each of the seven individual LXD infrared data sets. Although the subsolar latitude was within five degrees of the model's equator during all of our IRTF observations, the sub-observer latitude decreased substantially. Initially, we viewed ET70 from the ``top'' near the northern pole on February~11, but as time progressed, the view changed until we saw primarily the equator by February~21. That difference in viewing geometry is reflected in the thermal models and IR spectra from each subset as shown in \Fig{fig_therm_Feb11} through \Fig{fig_therm_Feb21}.
The left panels in each figure show the range of thermal models for the given subset, illustrating that there are a family of curves in single-scattering albedo and thermal inertia for a given choice of crater fraction, i.e., as we eliminate the beaming parameter and use the actual shape and spin state to investigate the thermal emission, there are multiple solutions that are equally good. As time progresses from \Fig{fig_therm_Feb11} to \Fig{fig_therm_Feb21}, the allowed family of curves shifts smoothly (downward and to the left) and grows smaller. This clearly demonstrates that the allowed models for the four subsets are not the same and that something is changing across the surface.

The right panels in \Fig{fig_therm_Feb11} through \Fig{fig_therm_Feb21} compare the measured thermal spectra for each subset to spectra from three models chosen to span the arcs of solutions. Moving from \Fig{fig_therm_Feb11} to \Fig{fig_therm_Feb21}, the thermal model parameters that provide good fits to the spectra from one date do not fit well for other dates. This is evidence of a smooth variation in the surface parameters between the northern latitudes and more equatorial ones. Whether it is a change in albedo, thermal inertia, surface roughness, or some combination, we cannot specifically say, but we can quantify the levels of the variations that are needed through comparisons such as these.

For instance, a model with $w = 0.10$, $f_c~=~0.2$, and $\Gamma~=~70$~J~m\textsuperscript{-2}~K\textsuperscript{-1}~s\textsuperscript{-1/2} provided a good fit to the February~21 data, but its model LXD spectra (2.2~to~4.1~microns) were too hot (too bright at long wavelengths) for the other two nights.
We examined the level of inhomogeneity that would be needed to match our observations by searching for the thermal parameters that would provide the best fits to each individual night's data. Based on allowing one parameter to vary while keeping the others fixed, we found that the first two nights' LXD spectra could be fit by decreasing the crater fraction from 0.2 to 0.1, or by increasing the thermal inertia from 70 to 100~J~m\textsuperscript{-2}~K\textsuperscript{-1}~s\textsuperscript{-1/2}, as illustrated in \Fig{fig_therm_params_lines}. The first two nights' LXD spectra could also be fit by increasing $w$ from 0.100 to 0.112.

The thermal model shown in \Fig{fig_tsky_colorized} ($w~=~0.108$, $f_c~=~0.2$, and $\Gamma~=~70$~J~m\textsuperscript{-2}~K\textsuperscript{-1}~s\textsuperscript{-1/2}) is slightly too hot at some observation times and too cold (too faint at long wavelengths) at other times. Specifically, the first three model LXD spectra for February~18 are slightly too hot, and the model LXD spectra for February~21 are too cold. In order to get good fits to the spectra from each night, the model's single-scattering albedo must be allowed to vary from 0.100~to~0.112. Figures S41 through S51 in the Supplementary material show spectra of three models with $f_c~=~0.2$ and $\Gamma~=~70$~J~m\textsuperscript{-2}~K\textsuperscript{-1}~s\textsuperscript{-1/2} that span that range of $w$.

\clearpage
\subsection{Comparisons with NEATM \label{sec_thermal_neatm}}

To determine whether simpler thermal models could provide an acceptable fit to our IRTF spectra, we compared our shape-based thermophysical models to thermal models using a sphere with negligible thermal inertia. We used a model that is similar to the widely used near-Earth asteroid thermal model (NEATM) described by \citen{Harris1998}, in which the albedo and beaming parameter are varied to match the observations. \citen{Howell2017} describe our NEATM-like modeling program in more detail.

A limitation of this simpler model is that the input reflectance curve cannot be as easily specified as it can for SHERMAN. An object with a strongly red-sloped reflectance spectrum like ET70 is thus more difficult to fit with a model that assumes a flatter input curve. To compare with the NEATM-like model spectra, we chose to normalize the spectra at 1.6~microns where the thermal contribution is negligible. The best-fit model was chosen based on the chi-squared value of the observed-model relative reflectance values in the 0.8-4.05~micron spectral region, but weighted more heavily towards the 3-4~micron region where the thermal contribution is greatest.

We ran over two thousand models covering a wide range of geometric albedo and beaming parameter values. Similar to our results using the more complicated thermophysical models with the ET70 shape, we found that no single set of NEATM parameters could provide an adequate fit to the IRTF spectra from all three nights. Furthermore, the LXD spectra from February~18 differ by enough that no single set of model parameters can fit all LXD spectra from that night. We note that although similar results are found here, the advantage of using our more detailed model over these NEATM-like models is that by explicitly taking the shape and illumination into account, the resulting thermal parameters are more physically based and the relationships among the parameters can be meaningfully explored (see \Fig{fig_therm_Feb11} through \Fig{fig_therm_Feb21}).

For NEATM-like thermal models, the best-fit beaming parameter ($\eta$) tends to increase as phase angle ($\alpha$) increases. For instance, \citen{Trilling2016} found a relation $\eta = 0.87 + (0.01$~deg$^{-1})\alpha$. Using assumed beaming parameters from that relation for each night's thermal models, we found that no single albedo value could fit all of the IRTF observations. The albedo would have to vary from 0.060 to 0.070, as shown in \Tab{tab_neatm}.

\begin{table}
\begin{tabular}{|lccc|}
\hline
Spectrum & $\alpha$ & Assumed $\eta(\alpha)$ & Best-fit albedo \\\hline
Feb-11 LXD   & $71^{\circ}$ & 1.58 & 0.070 \\
Feb-18 LXD B & $47^{\circ}$ & 1.34 & 0.070 \\
Feb-18 LXD D & $47^{\circ}$ & 1.34 & 0.060 \\
Feb-21 LXD A & $41^{\circ}$ & 1.28 & 0.065 \\\hline
\end{tabular}
\caption{
Best-fit albedo values for NEATM-like models with beaming parameters fixed at the values expected from the phase relation of \citen{Trilling2016}. Although these models are the best fits for their specific values of the beaming parameter, some of the fits are quite poor. For instance, the Feb-11 LXD model is $3 \sigma$ below the data (too cold) at 3~microns and $2.5 \sigma$ above the data (too hot) at 4~microns. Smaller values of the beaming parameter would give better fits for the earlier dates.
% For instance, the Feb-18 LXD~B model is at least two sigmas below the data (too cold) at 3~microns and at least two sigmas above the data (too hot) at 4~microns. Smaller values of the beaming parameter would give better fits for the earlier dates.
}
\label{tab_neatm}
\end{table}

%\clearpage
\section{Conclusions \label{sec_conc}}

We present an improved shape model and spin state for (162421) 2000~ET70 compared to that of \citen{Naidu2013}. Using both radar and lightcurve data, we found that the period-pole degeneracy allows for a sidereal rotation period of $8.944^{+0.013}_{-0.018}$~hours and a twenty-three~degree long arc of pole positions (\Fig{fig_banana_period}), a range that already excludes the pole of Naidu et al., which was based solely on radar data. Using our infrared observations, we limited the arc further to an angular length of about twelve~degrees, and we determined a best-fit pole at $(52^{\circ}, -60^{\circ}) \pm 6^{\circ}$ with a rotation period of $8.9444^{+0.0100}_{-0.0081}$~hours.
There will be opportunities to obtain photometry of ET70 in March of 2023 (expected $m_V \approx 18$) and in February of 2024 ($m_V \approx 17$), and additional observations could provide tighter constraints in the future. However, the next time that ET70 comes within 0.1~au of Earth will be in 2047, so no additional high-SNR radar observations of ET70 will be possible until then.

After extensive thermal modeling using our improved shape model and spin state, we found that our three nights of infrared observations could not be fit with a single homogeneous model. Instead, different thermal models were required for four different subsets of the data, and for any given subset there is a family of models, all of which points to variations in the thermal parameters across the surface of ET70. Similarly, \citen{Crowell2016} observed 1627~Ivar from the IRTF on multiple nights and found that no single set of thermal parameters could fit all of the observations; Ivar's surface properties also seem to be heterogeneous.

The ET70 data are not sufficient to allow a detailed determination of the thermal parameter values, but they do constrain the levels of inhomogeneity and suggest that it changes smoothly from northern latitudes down to equatorial latitudes. The later LXD spectra require thermal models whose parameters make the visible regions hotter, either with lower albedo, lower thermal inertia, greater roughness, or some combination.

The thermal models favor a surface that is smooth at sub-facet scales (i.e., lower values of crater fraction). \citen{Naidu2013} noted that ET70's relatively low radar circular polarization ratio, $\mu_C = 0.21 \pm 0.02$, implies a surface that is smoother than most NEAs at 10~cm scales. ET70's circular polarization ratio is less than the \citen{Benner2008} mean of 0.34 for all NEAs and consistent with the mean of 0.19 for P- and D-type asteroids. Based on the results from the thermal models, ET70's crater coverage fraction is probably less than 0.4 (assuming an opening angle $\gamma = 150^{\circ}$). This means that ET70's RMS slope $\rho$ is likely less than about 0.6 (which gives $\theta_{rms} = \arctan \rho \lesssim 30^{\circ}$). Both the radar observations and the infrared spectra suggest that ET70 has a fairly smooth surface, but those two types of observations are not necessarily probing the surface roughness at the same spatial scales.

Only small variations in the thermal properties are needed to match the spectra in each infrared spectral subset: a change of about 0.01 in albedo, a difference of about 0.1 in crater coverage fraction, or a change in the thermal inertia on the level of 50~J~m\textsuperscript{-2}~K\textsuperscript{-1}~s\textsuperscript{-1/2}. Such variations are physically reasonable; for example, the albedo variations needed for ET70 are similar to the relative contrasts of $\sim$6\% seen by NEAR on C-type asteroid 253~Mathilde \cite{Clark1999}.
Models at the lower end of our possible albedos (\Fig{fig_therm_Feb11} through \Fig{fig_therm_Feb21}) allow for a thermal inertia that is compatible with the \citen{Delbo2007} average for kilometer-sized near-Earth asteroids, $200~\pm~40$~J~m\textsuperscript{-2}~K\textsuperscript{-1}~s\textsuperscript{-1/2}. However, there are also possible models with considerably lower thermal inertias -- perhaps closer to the \citen{Capria2014} average thermal inertia of Vesta, $30~\pm~10$~J~m\textsuperscript{-2}~K\textsuperscript{-1}~s\textsuperscript{-1/2}.

Based on ET70's size and its absolute magnitude, \citen{Naidu2013} noted that ET70 must have either a very low albedo or a strange phase function. However, the previously reported values for ET70's absolute visual magnitude were extrapolated from high phase angles or had large uncertainties. For ET70, the Minor Planet Center assumes a phase slope of $G = 0.15$ and gives an absolute visual magnitude of $H_V = 18.0$ \cite{MPO294169}. However, that absolute magnitude is based on observations that were all taken at phase angles of over 40~degrees, with large uncertainties in the observed magnitudes. Using $H_V = 18.0$ with our value $D = 2.10$~km (from shape modeling), the standard relation $D \sqrt{p_V} = K \times 10^{-H_V/5}$, where $K = 1329$~km \cite{Pravec2007}, yields a geometric albedo of $p_V = 0.025$, which is not consistent with the thermal models. With $G = 0.17$, $p_V = 0.025$ corresponds to $w = 0.03$ \cite{Verbiscer1995}.

We do not have absolute photometry that can provide an independent estimate of ET70's absolute magnitude, but we can calculate its absolute magnitude from our values of the radar-derived size and the thermal model's albedo. A conservative estimate of the allowed range for the thermal models' single-scattering albedo is 0.06~to~0.15. With $G = 0.17$, that range of single-scattering albedo converts to geometric albedo $p_V \approx 0.09 \pm 0.04$ \cite{Verbiscer1995}. Taking $D = 2.10 \pm 0.07$~km and $p_V = 0.09 \pm 0.04$, ET70's absolute magnitude is $H_V = 16.6 \pm 0.5$, with the uncertainty in ET70's albedo having the greatest contribution to the uncertainty in $H_V$.

$H_V \approx 16.6$, combined with the Minor Planet Center's tabulated magnitudes, would imply a fairly strong opposition effect, with a phase slope $G \lesssim 0$. However, given that the MPC magnitudes have large uncertainties, and that our thermal models are not very sensitive to the value of $G$, the available data are not sufficient to determine $G$.

Our results imply that even small NEAs are complex geologic objects with inhomogeneous surface properties. The thermal parameters derived from observations on a single night may depend more on the local surface properties than generally assumed, and thus may not be representative of the asteroid surface as a whole. Although we cannot uniquely determine the thermal parameters for ET70 from our dataset as the number of observations is too sparse, by using a realistic representation of the shape and the actual spin state, we have been able to investigate the surface of ET70 in terms of physically meaningful thermal parameters. This approach opens the door to treating NEAs as truly physical objects, complete with variations in surface thermal parameters that we can constrain and explore from ground-based observations at multiple viewing geometries.

\section{Acknowledgments \label{sec_ack}}

This project was partially supported by NASA grants NNX10AP87G, NNX12AF24G, and NNX13AQ46G; and by NSF grant AST-1109855. The first author was supported by a NASA Earth and Space Science Fellowship (NASA grant NNX15AR14H).
Several co-authors were Visiting Astronomers at the InfraRed Telescope Facility, which is operated by the University of Hawaii under contract NNH14CK55B with the National Aeronautics and Space Administration.
The authors thank the staff members at Arecibo Observatory, at the InfraRed Telescope Facility, and at Palomar Observatory for assistance with the observations.
The Arecibo Observatory is operated by SRI International in partnership with Ana G. M\'{e}ndez -- Universidad Metropolitana and the Universities Space Research Association, under a cooperative agreement with the National Science Foundation (AST-1100968). The Arecibo Planetary Radar program is supported by NASA's Near Earth Object Observation program.
The authors also thank the two reviewers for their constructive feedback.
Taxonomic type results presented in this work were determined using a Bus-DeMeo Taxonomy Classification Web tool by Stephen M. Slivan, developed at MIT with the support of National Science Foundation Grant 0506716 and NASA Grant NAG5-12355.
This research has made use of NASA's Astrophysics Data System Bibliographic Services.
The authors gratefully acknowledge the teams that produce the following free software packages, which were used for this work: Gnuplot\footnote{\url{http://www.gnuplot.info/}}, ImageMagick\footnote{\url{https://imagemagick.org/}}, Kile\footnote{\url{http://kile.sourceforge.net/}}, LibreOffice\footnote{\url{https://www.libreoffice.org/}}, Perl\footnote{\url{https://www.perl.org/}}, Python\footnote{\url{https://www.python.org/}}, IPython \cite{Perez2007}, Matplotlib \cite{Hunter2007}, NumPy \cite{vanderWalt2011}, Pillow\footnote{\url{https://python-pillow.org/}}, and SciPy\footnote{\url{http://www.scipy.org/}}.

\section{Supplementary material \label{sec_sup}}
%Supplementary figures associated with this article can be found in the online version.
Supplementary figures associated with this article can be found at \url{http://astro.cornell.edu/~seanm/2000et70/}.
%, at [URL].
% TODO: Give URL, once a DOI has been assigned!

\section{References \label{sec_refs}}

\bibliographystyle{model2-names}\biboptions{authoryear}
\bibliography{ET70_v22}
%\bibliography{ET70_v22.bib}
%\addbibresource{ET70_v22.bib}

\end{document}